\documentclass[prl,preprint]{revtex4}
\pdfoutput=1
\usepackage[utf8]{inputenc}
\usepackage{amsmath,amssymb}
\usepackage{graphicx}
\usepackage{textcomp}
\setcitestyle{super}

\begin{document}
\title{Label-free Plasmonic Detection of Untethered Nanometer-sized Brownian Particles}
\author{Martin Dieter Baaske}
\author{Peter Sebastian Neu}
\author{Michel Orrit*}
\affiliation{Huygens-Kamerlingh Onnes Laboratory, Leiden University, Postbus 9504, 2300 RA Leiden, The Netherlands \\ *email: orrit@physics.leidenuniv.nl }
\date{\today}

\maketitle

\textbf{Optical detection of individual nanometer-sized analytes, virus particles, and protein molecules holds great promise for understanding and control of biological samples and healthcare applications. As fluorescent labels impose restrictions on detection bandwidth and require lengthy and invasive processes, label-free optical techniques are highly desirable. Powerful label-free optical methods have recently emerged, such as interferometric scattering microscopy\cite{Piliarik2014,Liebel2017,Huang2017,McDonald2018,Young2018}, plasmonic nanoparticle-based assays \cite{Zijlstra_2012,Ament2012,Rosman2013,Beuwer2015,Ye2018} and microcavity-based assays \cite{Vollmer2008,Zhu2009,He2011,Dantham2013,Bei-Bei2014,Baaske2014,Kim2016,Baaske2016,Chen2017,Kim2017}. Although highly sensitive, these methods are so far restricted to integration times in excess of microseconds. This often imposes a requirement to impede analyte motion during these periods via specific molecular tethers, unspecific adsorption or confining arrangements\cite{Ignatovich2006a,Krishnan2010,Faez2015,Squire2019}.
Here we introduce an optical technique capable of transforming gold nanorods commonly used as photostable labels\cite{Wang2005,Huang2006,Durr2007} into highly localized high-speed probes. Our method provides a time resolution well below microseconds. This mitigates the requirement for molecular tethers and allows us to detect single untethered nanoparticles in Brownian motion traversing sub atto-liter sensing volumes. Our method opens a novel gateway for the investigation of highly localized and highly dynamic nanoscale systems and constitutes a  first step towards the label-free recognition of single untethered proteins.}

Label-free optical techniques so far regularly rely on chemical tethers or other means that impede the analyte's motion. These chemical tethers or receptor molecules fulfill a dual purpose. Firstly they provide the specificity, i.e., they ideally only interact with one species of target molecules and thereby provide selective identification. Their second purpose is to hold the target analyte fast for time periods long enough to enable detection. The requirement for chemical specificity can be relaxed to a large extent if the assay directly or indirectly measures several physical properties of the analyte, i.e., its charge, mass or polarizability. Then, unspecific adsorption to a surface is sufficient\cite{Young2018}. For plasmonic nanoparticle-based assays permanent or long adsorption duration are undesirable as their limited surface area allows only few analytes to bind. Highly specific chemical interaction, however, usually entails high affinity (i.e., strong bonds) - making the multiplexed read out of many particles a necessity \cite{Rosman2013,Beuwer2015}.
Consequently, a nanoparticle-based method which would lift the requirement for specific chemical modifications would be highly advantageous but requires accurate determination of more physical properties, such as Stokes radius or charge, to enable identification. 
However, to provide access to these properties, such a sensor must probe sub-attoliter volumes and therefore needs to be fast. The current state of the art for on-the-fly plasmonic detection was established by Wulf et al.\ who detected diffusing particles as they propagate through a sensing nanorod's near field \cite{Wulf2016}. Their method required tracking the whole spectrum of the rod's plasmon resonance and therefore was limited to $\approx0.1$ millisecond time resolution and comparatively large particles (diameter$\geq20$\,nm). In the following we will demonstrate an optical method that improves the time resolution $10^4$-fold, the sensitivity with respect to analyte polarizability more than $100$-fold and in consequence is capable of recognizing even single analytes.

In order to resolve such short-lived and minute intensity perturbations due to shifts of a gold nanorod's (GNR) longitudinal surface plasmon resonance (LSPR, frequency $\nu_0$, half width at half maximum $\Gamma$), one must overcome fundamental noise sources (essentially photon shot noise) as well as experimental noise from laser, detector, and residual vibrations and drifts of the setup \cite{Jollans2019}. To do so, one must optimize the signal-to-noise ratio for fluctuations in detected power caused by shifts of the GNR's plasmon resonance. 
\begin{figure}[htbp]
    \centering
    \includegraphics{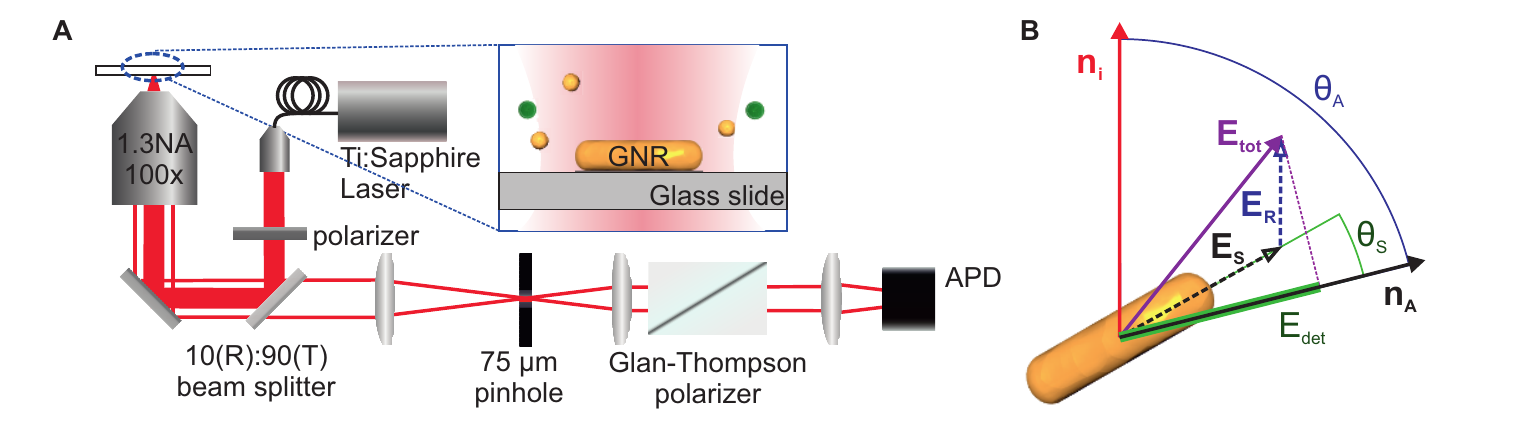}
    \caption{Experimental method: \textbf{A} shows the setup used to monitor minute fluctuations of scattered intensity due to perturbations of a GNR's near field (inset) by freely diffusing nanoscopic analytes. \textbf{B} shows how the ratio of scattered $\mathbf{E}_S$ to reflected field $\mathbf{E}_r$ components contributing to $E_{det}$ can be chosen by varying the incident field polarisation $\mathbf{n}_i$ and the analyser (Glan-Thompson) $\mathbf{n}_A$ orientation i.\ e.\ the angles $\theta_A$ and $\theta_S$.
    }
    \label{fig:1}
\end{figure}
We do this via the confocal microscopy setup shown in Fig.\,1A, which allows us to utilize the intrinsic scattering anisotropy of GNRs. As depicted in Fig.\,1B this in principle allows us to choose the ratio of reflected $E_R$ over scattered field $E_S$ contributions to the detected field component $E_{det}$ for an arbitrary GNR orientation and therefore tune the signal to noise ratio for individual NRs (see suppl. information section 1). 
In order to avoid contributions from linearly birefringent components, such as dielectric mirrors and beam-splitters, to the analyzed signals we, however, choose to restrict our proof of principle measurements to close to cross-polarized configurations. While this does not yield the optimal signal-to-noise ratio it rejects contributions from isotropic scatters entering the confocal volume thus allowing to unequivocally attribute changes in intensity to changes in the scattering cross sections of the monitored NR.
Another parameter that has to be considered is the temperature increase of the NRs for which we find $\Delta T\lesssim 8$\,K under our experimental conditions (suppl. section S2).

We record continuous $0.1$\,ms traces of scattered intensity: $\tilde{I}(t)=I(t)-\langle I\rangle$, from which we compute the respective normalized autocorrelation curves
\begin{equation}
G(\tau)=\frac{\langle \tilde{I}(t+\tau)\tilde{I}(t)\rangle }{\langle \tilde{I}^2 \rangle}
\end{equation} 
using the Wiener–Khinchin theorem in order to obtain ensemble properties. Note that we normalize to the variance of the intensity fluctuations instead of the squared average intensity, because we use an AC-coupled detector in order to reject low-frequency noise. 
In addition to autocorrelation curves from single traces, we also discuss averages $G_N(\tau)$ over $N$ traces.
We find that all analytes discussed in the following exhibit autocorrelation curves that are fitted well with stretched exponentials:
\begin{equation}
    G_{fit}(\tau)=Ae^{-(\tau/\tau_D)^\beta}
\end{equation}
where $A$ denote the amplitude, $\beta$ the stretch-exponent and $\tau_D$ the decay time. 

In order to show that our sensor can obtain information in a relatively crowded environment i.e. at analyte concentrations in excess of $0.1$\,mM, which are usually not accessible to image-based methods, we have prepared a microemulsion of oil in water. Microemulsions are stable physical phases of ternary surfactant-oil-water mixtures and do not suffer from the drawbacks of unspecific sticking that are commonly encountered for proteins.

This way we circumvent the need for chemical surface modifications in these pilot experiments. Specifically we have chosen a nonionic microemulsion system consisting of a soybean oil/ polyoxyethylene-10-oleyl ether (Brij-O10)/ water mixture\cite{Warisnoicharoen2000,WARISNOICHAROEN2000_2} ($4 \% / 16\% / 80\% $) that forms stable and mono-disperse micelles, or nanodroplets with $(8.1\pm2.6)$\,nm diameter, as determined via dynamic light scattering (see suppl. Fig.\,S4). As micelles have a refractive index of $1.48$\cite{Malcolmson2002}, they mimic proteins of $\approx 250$\,kDa molecular weight in size, shape, and polarizability. The concentration of these solutions is typically $\approx 1$\,mM i.e. $20\%$ volume fraction (oil + detergent).
In this high concentration regime the unequivocal recognition of single analytes is not possible as more than one analyte particle will regularly be present in the NR's near field. Nonetheless ensemble properties can be obtained from intensity trace autocorrelations.
\begin{figure}[hb]
    \centering
    \includegraphics{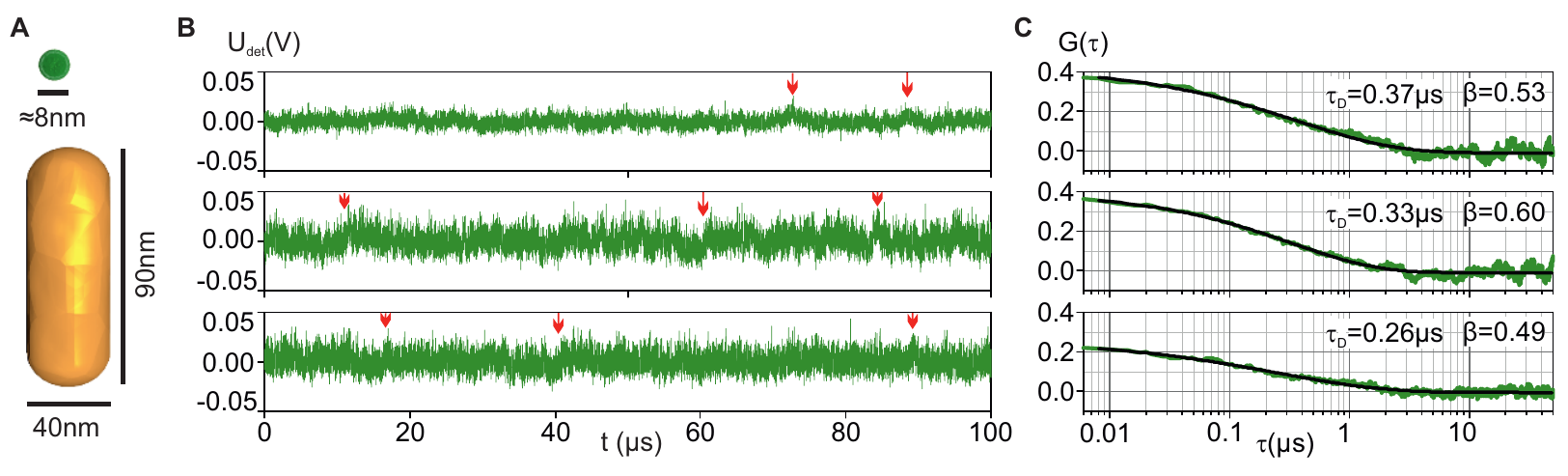}
    \caption{Concentration fluctuations in a microemulsion: The dimensions of the sensor rod and the microemulsion nanodroplets simulating $250$\,kDa proteins are depicted in \textbf{A}. Panel \textbf{B} shows $0.1$\,ms long intensity traces which exhibit sub-microsecond variations of the scattered intensity caused by number fluctuations of nanodroplets (examples indicated by red arrows). The three traces were recorded on three different rods.
     Panel \textbf{C} shows the corresponding autocorrelation curves (green) together with their stretched exponential fits (black). 
    \label{fig:2}}
\end{figure}

The intensity traces shown in Fig.\,2B exhibit clear perturbations of (sub-)microsecond duration caused by these microemulsion nanodroplets - the positive and negative signs of these events and their high rate of occurrence suggest that these cannot be unequivocally contributed by single micelles, but rather number fluctuations of nanodroplets in the near field. Even autocorrelation curves of $0.1$\,ms intervals exhibit significant contrast despite sub-microsecond relaxation times (see Fig.\,2C). Averaged autocorrelations ($G_{1000}$) obtained from different gold nanorods in different microemulsion samples show only minor deviations with respect to $\beta$. The $\tau_D$ times, however, range from $0.25$ to $1.0$\,\textmu s  (compare also Suppl. Fig. S4). 
These differences in $\tau_D$ and $\beta$ might well reflect individual differences in size and shape of the single GNRs and their respective near field distributions.
The diffusion length $L=\sqrt{2D\tau_D}$, where the diffusivity $D$ is given by the Stokes Einstein relation, associated to the determined $\tau_D$ values falls into the interval $L=(4-12)$\,nm. Here the higher value is comparable to near field decay length\cite{Wulf2016}, whereas the lower value matches the analyte's radius. The latter case might indicate that the nanodroplets can probe point-like defects on the GNR's surface that are associated with high field strength in their proximity\cite{Dantham2013,Baaske2014,Benz2016}.   
We also find the amplitude of our autocorrelation curves to decrease with nanodroplet concentration as we step-wise dilute the sample to a quarter of its initial concentration $C_0$ (see suppl. Fig. S4B). This reflects the reduction in the number of nanodroplets perturbing the GNR's near field in a given time interval. We want to note here that usually the autocorrelation's contrast can not be unequivocally attributed to either the magnitude of single analyte perturbations  or the rate at which these occur (suppl. sect. S4).  

In order to investigate our sensor's performance in the sub-micromolar concentration regime we utilize citrate capped gold nanoparticles (GNPs) with $5$\,nm diameter and concentrations $<50$\,nM. 
\begin{figure}[htbp]
    \centering
    \includegraphics{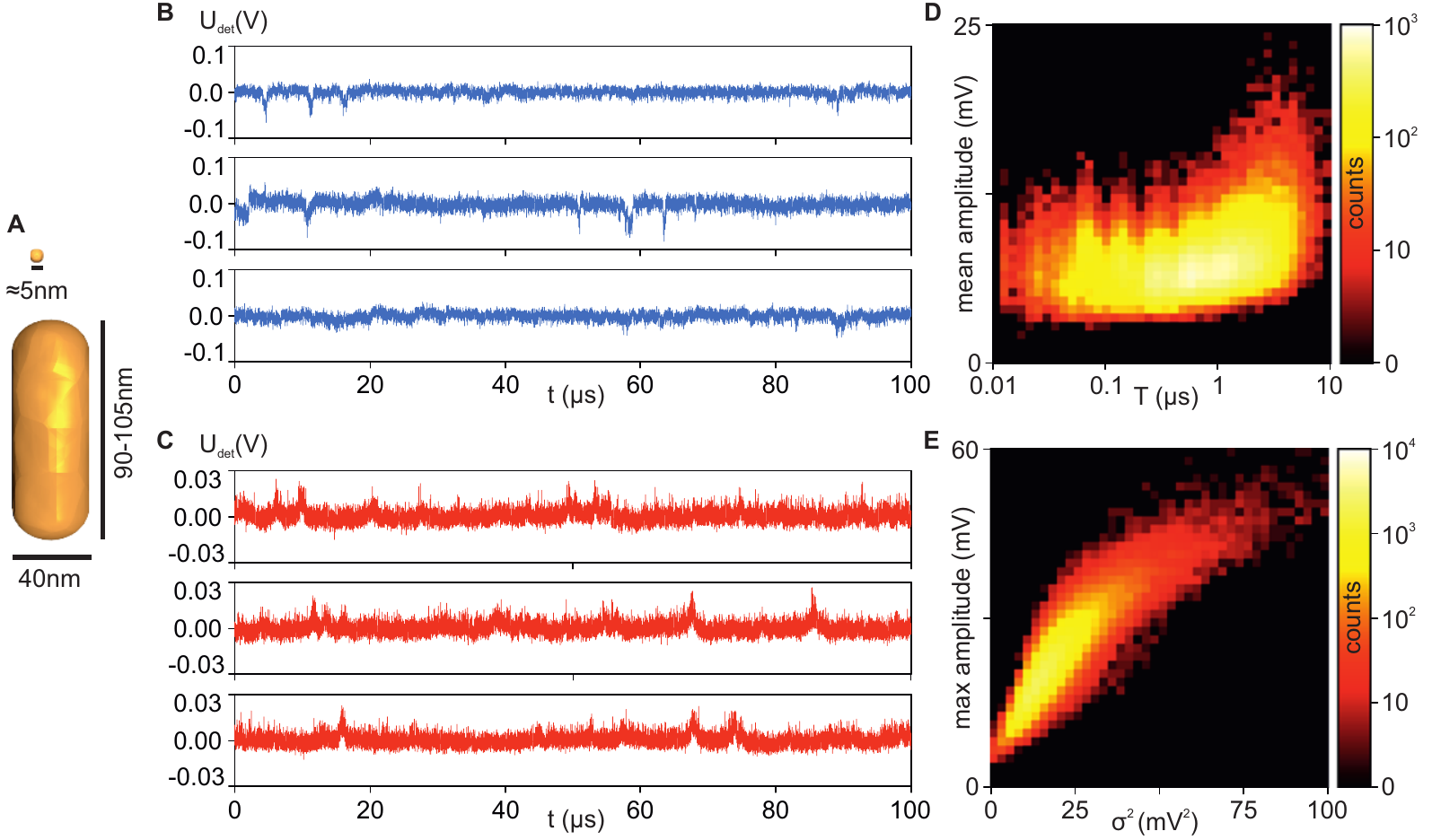}
    \caption{Single-particle detection: Panel \textbf{A} shows the dimensions of the nanorod sensors and the analyte nanoparticles. Example intensity traces exhibiting clear intensity bursts caused by single analyte GNPs are shown in panels \textbf{B} and \textbf{C}. Burst are towards lower \textbf{B} and higher \textbf{C} intensities for the two different NRs (\textbf{B}: $125\times40$\,nm  and \textbf{C}: $90\times40$\,nm) probed on the short(\textbf{B}) and long (\textbf{C}) wavelength side of their LSPRs, respectively. Panels \textbf{D} and \textbf{E} show distributions of event mean amplitude vs. event duration $T$ and maximum amplitude vs. variance ($\sigma^2$), respectively (same NR as \textbf{C}). Both distributions exhibit clear correlations between the corresponding properties.}
    \label{fig:3}
\end{figure}
In this regime we recognize clear spike/burst-like intensity perturbations due to single particles (Fig. 3B,C). We have developed an algorithm that recognizes these perturbations (suppl. Sect. S6). 
In general we find that these perturbations show the expected sign dependent on which side of a nanorod's LSPR is probed. Specifically this means an intensity decrease (increase) for NRs probed on the short (long) wavelength side of their LSPRs as analytes with positive excess polarizabilities entering the NRs near field will cause the resonance to shift towards longer wavelengths.
We further find that the distributions of inter-event durations follow poissonian statistics (see Suppl. section S7). Note that the respective autocorrelations and stretch exponential fits for the example traces in Fig.\,3 are shown in the Supplementary Information (Fig.\,S10). In contrast to these autocorrelations the data extracted from single particle events provides access to additional dimensions for analysis i.e. additional means to discern between analytes and their diffusive behavior with respect to experimental conditions. As shown in figure 3D we find clear correlations between the mean amplitudes and the durations ($T$) of individual events. The mean amplitude of an event is a measure for the average integrated field strength an individual particles sees along its trajectory through the detection volume. The duration $T$ provides a measure for the time a particle spent inside the detection volume without interruption.
Specifically, we find that particles which remain inside the NR's detection volume for longer are also more likely to possess trajectories with higher integrated field strength yielding higher mean amplitudes. This reflects that particles traversing the near field for longer are also more likely to penetrate deeper into the NR's near field towards its tips where the near field is the strongest. Trajectories that bring particles close to the NR's tips have to traverse the  detection volume at least twice and thus require longer minimum durations as compared to trajectories which just graze the detection volume's outer boundary. 
We also find a clear correlation between the maximum amplitudes and the variance of individual events (compare Fig.\,3E). The maximum amplitude provides a measure for the highest field strength a particle sees along its trajectory i.e. a proxy for how close an individual particle came to the NR's surface and especially it's tips. The variance 
\begin{equation} 
\sigma^2= \frac{1}{T}\sum_T \left( I(t)-\langle I \rangle\right)^2
\end{equation}
is a measure for how strong the amplitude $I$ of an individual event fluctuates about its mean $\langle I \rangle$ throughout its duration $T$ and therefore provides a proxy for the overall variation of the field's strength along a particle's trajectory. 
The NR's near field decays non-linearly and rapidly with increasing distance from the NR's tips. Thus we expect that particles with trajectories that enter zones with higher field strength need to diffuse through zones that exhibit stronger field gradients. Indeed we find this correlation reflected in our data as events with higher maximum amplitudes exhibit higher variance (compare Fig.\,3E).
Our simulation results (see suppl. section S8) indicate that in future studies this type of analysis can be used to discern between different analytes possibly even on a single shot basis. 
This type of discrimination is not directly possible from autocorrelations or via ensemble measurements like traditional dynamic light scattering without prior knowledge of analyte composition. We also want to point out that the single event analysis can in principle be extended to include higher statistical moments like kurtosis and skewness hence providing additional dimensions for the discrimination of analytes (see Fig. S11).

The measurements shown in Fig. 3 were performed at an ionic strength of $50$\,mM (3B) and $120$\,mM (3C, 3D and 3E), respectively.
Both the citrate-capped GNP and the nanorods are negatively charged at neutral pH and thus repel each other. The range of this repulsive interaction can be altered via the solution's ionic strength, i.e., Debye-screening (suppl. section S5)\cite{Israelachvili2011}. 
In fact we do not recognize any events in the presence of GNPs without additional electrolyte in the solution (Suppl. Section. S6). 
\begin{figure}[hbt]
    \centering
    \includegraphics{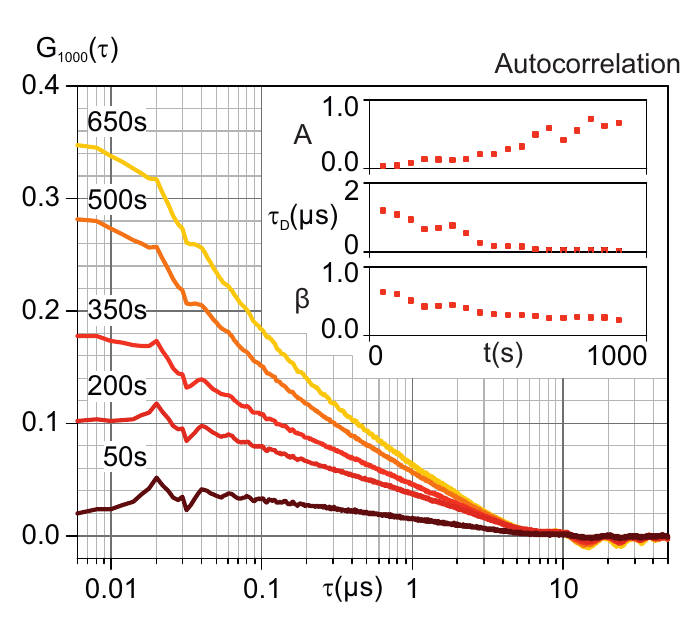}
    \caption{Gold nanoparticle detection at varying ionic strength: Panel \textbf{A} shows the change of averaged trace autocorrelation over time upon injection of sodium chloride (final concentration $30$\,mM) alongside the stretch exponential fit parameters (inset).}
    \label{fig:4}
\end{figure}
To further investigate the influence of Debye screening on the diffusion of the $5$\,nm diameter GNPs, we record intensity traces upon the injection of sodium chloride to a final concentration of $30$\,mM into Milli-Q water premixed with $8$\,nM GNPs. Following the injection the local ionic strength around the NR will increase over time until it reaches equilibrium. Due to this process the volume in which the repulsive Coulomb interaction between analyte GNPs and the sensor nanorod can dominate the Brownian motion will decrease over time, allowing the analytes to come ever closer to the NRs surface.
This process is reflected in our intensity autocorrelations (Fig.\,4) as an increase in contrast over time, shorter $\tau_D$ values and lower stretch exponents $\beta$. As mentioned above discrimination between contributions from perturbation rate and perturbation magnitude is not directly possible via autocorrelation contrast alone. 
Our sensor's capacity to resolve perturbations by single particles, however, provides not only direct access to event rates and amplitudes but also allows us to correlate statistical moments (Fig.\,5). 
\begin{figure}[hbt]
    \centering
    \includegraphics{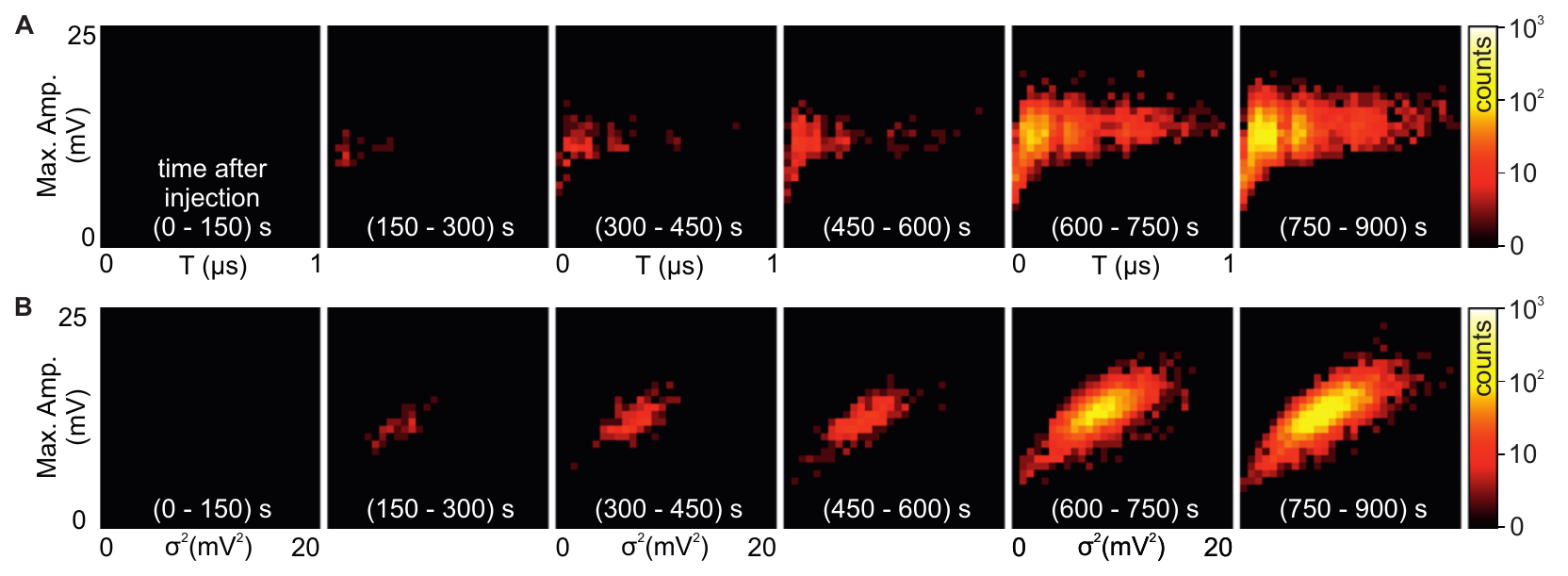}
    \caption{Single gold nanoparticle detection at varying ionic strength: Both panels represent the change of single-particle event properties over time upon injection of sodium chloride (final concentration $30$\,mM). \textbf{A}: Distribution of maximum amplitudes vs. event duration. \textbf{B}: Distribution of maximum amplitudes vs. variance ($\sigma^2$). Based on the same dataset as Fig.\,4.}
    \label{fig:5}
\end{figure}
We find that as the ionic strength increases with time firstly the number of events increases (Fig.\,5 $\leq 600$\,s) whereas event durations are short, maximum amplitudes weak event variance is overall low. This indicates that the analyte particles are at first only gaining access to growing zones of relatively low field strength and thus weak field gradients namely the outer layers of the NR's near field. Starting $600$\,s after the injection the maximum amplitudes and especially the duration of the single particle perturbations increase significantly towards the end of the measurement (Fig.\,5A). This shows that the analyte particles can now penetrate significantly deeper into the NR's near field and access zones with higher field strength. The clear correlation between maximum amplitudes and the variance (see Fig.\,5B) of individual events, as found previously (compare Fig.\,3E), further supports this interpretation.
We think our finding that event duration and amplitude start to increase at higher ionic strength is related to the correlation of local surface charge and local near field strength. Both are highest on the NR's tips. In consequence access towards the NR's tips might require a higher local ionic strength than access towards the sides of the NR and thus occurs later in our measurement.
Nonetheless the overall low maximum amplitudes, durations and the absence of the additional sharp tail in the maximum amplitude vs. variance distribution as found at higher ionic strength (compare Fig.\,3E) suggest that the observed NR's tips were not fully accessible at an ionic strength of $30$\,nM.
This type of study again shows that our sensor is capable of extracting information beyond the level commonly accessible via ensemble based measurements and moreover demonstrates our sensor's potential for the future discrimination of analytes with varying charge - for example via the application of controlled electrostatic potentials throughout the sample cell.

In conclusion we have demonstrated the detection of sub-$10$\,nm particles as they undergo Brownian motion through the nanometer-sized near field of single gold nanorods with sub microsecond time resolution. Our sensor is capable to perform ensemble type measurements in concentration ranges up to $1$\,mM, which are not accessible by label-free image-based techniques.
The successful detection of microemulsion nanodroplets with dimensions and optical properties similar to those of $\approx250$\,kDa proteins shows the promise of our technique for the label-free detection and identification of biological analytes without specific chemical receptors. The clear bursts obtained from single gold particles further suggest that single protein molecule recognition will be possible with further optimization.
We further found that individual GNRs can exhibit different correlation functions for the same analyte solution. We think this reflects the variation in near field distributions of individual gold nanorods and shows the promise of our method for their characterization. Despite these differences, individual nanorods, once calibrated with a standard or an appropriate optical technique, such as enhanced fluorescence\cite{Caldarola2018}, may be used for the accurate sizing of analytes.
Furthermore, we have demonstrated that our system is capable of recognizing single analyte particles with $5$\,nm diameter at lower concentration levels of $\approx 10$\,nM. We have shown that this single particle resolution provides access to additional layers of information, which are not directly accessible via ensemble methods and hold promise for the future discrimination of particle properties possibly even on a single shot basis. 
In principle, our technique does not require the immobilization of the gold nanorods onto a glass surface. Any environment restricting rotational diffusion, e.g., a cell membrane or other fixed structures could be used to optimize the scattering signal of the nanorod.

\section{Acknowledgements}
This work was supported by the Netherlands Organisation for Scientific Research (NWO) and has received funding from the European Union’s Horizon 2020 research and innovation programme under the Marie Skłodowska-Curie Grant Agreement no. 792595 (MDB).

\section{Author Contributions:}
MDB and MO conceived the idea. MDB and PSN built the optical setup and performed experiments. MDB performed data analysis. MDB and MO wrote the manuscript. All authors commented on the manuscript. 

\bibliography{references}

\begin{thebibliography}{35}%
\makeatletter
\providecommand \@ifxundefined [1]{%
 \@ifx{#1\undefined}
}%
\providecommand \@ifnum [1]{%
 \ifnum #1\expandafter \@firstoftwo
 \else \expandafter \@secondoftwo
 \fi
}%
\providecommand \@ifx [1]{%
 \ifx #1\expandafter \@firstoftwo
 \else \expandafter \@secondoftwo
 \fi
}%
\providecommand \natexlab [1]{#1}%
\providecommand \enquote  [1]{``#1''}%
\providecommand \bibnamefont  [1]{#1}%
\providecommand \bibfnamefont [1]{#1}%
\providecommand \citenamefont [1]{#1}%
\providecommand \href@noop [0]{\@secondoftwo}%
\providecommand \href [0]{\begingroup \@sanitize@url \@href}%
\providecommand \@href[1]{\@@startlink{#1}\@@href}%
\providecommand \@@href[1]{\endgroup#1\@@endlink}%
\providecommand \@sanitize@url [0]{\catcode `\\12\catcode `\$12\catcode
  `\&12\catcode `\#12\catcode `\^12\catcode `\_12\catcode `\%12\relax}%
\providecommand \@@startlink[1]{}%
\providecommand \@@endlink[0]{}%
\providecommand \url  [0]{\begingroup\@sanitize@url \@url }%
\providecommand \@url [1]{\endgroup\@href {#1}{\urlprefix }}%
\providecommand \urlprefix  [0]{URL }%
\providecommand \Eprint [0]{\href }%
\providecommand \doibase [0]{https://doi.org/}%
\providecommand \selectlanguage [0]{\@gobble}%
\providecommand \bibinfo  [0]{\@secondoftwo}%
\providecommand \bibfield  [0]{\@secondoftwo}%
\providecommand \translation [1]{[#1]}%
\providecommand \BibitemOpen [0]{}%
\providecommand \bibitemStop [0]{}%
\providecommand \bibitemNoStop [0]{.\EOS\space}%
\providecommand \EOS [0]{\spacefactor3000\relax}%
\providecommand \BibitemShut  [1]{\csname bibitem#1\endcsname}%
\let\auto@bib@innerbib\@empty
\bibitem [{\citenamefont {Piliarik}\ and\ \citenamefont
  {Sandoghdar}(2014)}]{Piliarik2014}%
  \BibitemOpen
  \bibfield  {author} {\bibinfo {author} {\bibfnamefont {M.}~\bibnamefont
  {Piliarik}}\ and\ \bibinfo {author} {\bibfnamefont {V.}~\bibnamefont
  {Sandoghdar}},\ }\bibfield  {title} {\bibinfo {title} {{Direct Optical
  Sensing of Single Unlabelled Proteins and Super-Resolution Imaging of their
  Binding Sites}},\ }\href@noop {} {\bibfield  {journal} {\bibinfo  {journal}
  {Nat. Commun.}\ }\textbf {\bibinfo {volume} {5}},\ \bibinfo {pages} {4495}
  (\bibinfo {year} {2014})}\BibitemShut {NoStop}%
\bibitem [{\citenamefont {Liebel}\ \emph {et~al.}(2017)\citenamefont {Liebel},
  \citenamefont {Hugall},\ and\ \citenamefont {van Hulst}}]{Liebel2017}%
  \BibitemOpen
  \bibfield  {author} {\bibinfo {author} {\bibfnamefont {M.}~\bibnamefont
  {Liebel}}, \bibinfo {author} {\bibfnamefont {J.~T.}\ \bibnamefont {Hugall}},\
  and\ \bibinfo {author} {\bibfnamefont {N.~F.}\ \bibnamefont {van Hulst}},\
  }\bibfield  {title} {\bibinfo {title} {Ultrasensitive label-free nanosensing
  and high-speed tracking of single proteins},\ }\href@noop {} {\bibfield
  {journal} {\bibinfo  {journal} {Nano Lett.}\ }\textbf {\bibinfo {volume}
  {17}},\ \bibinfo {pages} {1277} (\bibinfo {year} {2017})}\BibitemShut
  {NoStop}%
\bibitem [{\citenamefont {Huang}\ \emph {et~al.}(2017)\citenamefont {Huang},
  \citenamefont {Zhuo}, \citenamefont {Chou}, \citenamefont {Lin},
  \citenamefont {Chang},\ and\ \citenamefont {Hsieh}}]{Huang2017}%
  \BibitemOpen
  \bibfield  {author} {\bibinfo {author} {\bibfnamefont {Y.-F.}\ \bibnamefont
  {Huang}}, \bibinfo {author} {\bibfnamefont {G.-Y.}\ \bibnamefont {Zhuo}},
  \bibinfo {author} {\bibfnamefont {C.-Y.}\ \bibnamefont {Chou}}, \bibinfo
  {author} {\bibfnamefont {C.-H.}\ \bibnamefont {Lin}}, \bibinfo {author}
  {\bibfnamefont {W.}~\bibnamefont {Chang}},\ and\ \bibinfo {author}
  {\bibfnamefont {C.-L.}\ \bibnamefont {Hsieh}},\ }\bibfield  {title} {\bibinfo
  {title} {Coherent brightfield microscopy provides the spatiotemporal
  resolution to study early stage viral infection in live cells},\ }\href
  {https://doi.org/10.1021/acsnano.6b05601} {\bibfield  {journal} {\bibinfo
  {journal} {ACS Nano}\ }\textbf {\bibinfo {volume} {11}},\ \bibinfo {pages}
  {2575} (\bibinfo {year} {2017})}\BibitemShut {NoStop}%
\bibitem [{\citenamefont {McDonald}\ \emph {et~al.}(2018)\citenamefont
  {McDonald}, \citenamefont {Gemeinhardt}, \citenamefont {K{\"{o}}nig},
  \citenamefont {Piliarik}, \citenamefont {Schaffer}, \citenamefont
  {V{\"{o}}lkl}, \citenamefont {Aigner}, \citenamefont {Mackensen},\ and\
  \citenamefont {Sandoghdar}}]{McDonald2018}%
  \BibitemOpen
  \bibfield  {author} {\bibinfo {author} {\bibfnamefont {M.~P.}\ \bibnamefont
  {McDonald}}, \bibinfo {author} {\bibfnamefont {A.}~\bibnamefont
  {Gemeinhardt}}, \bibinfo {author} {\bibfnamefont {K.}~\bibnamefont
  {K{\"{o}}nig}}, \bibinfo {author} {\bibfnamefont {M.}~\bibnamefont
  {Piliarik}}, \bibinfo {author} {\bibfnamefont {S.}~\bibnamefont {Schaffer}},
  \bibinfo {author} {\bibfnamefont {S.}~\bibnamefont {V{\"{o}}lkl}}, \bibinfo
  {author} {\bibfnamefont {M.}~\bibnamefont {Aigner}}, \bibinfo {author}
  {\bibfnamefont {A.}~\bibnamefont {Mackensen}},\ and\ \bibinfo {author}
  {\bibfnamefont {V.}~\bibnamefont {Sandoghdar}},\ }\bibfield  {title}
  {\bibinfo {title} {{Visualizing Single-Cell Secretion Dynamics with
  Single-Protein Sensitivity}},\ }\href
  {https://doi.org/10.1021/acs.nanolett.7b04494} {\bibfield  {journal}
  {\bibinfo  {journal} {Nano Lett.}\ }\textbf {\bibinfo {volume} {18}},\
  \bibinfo {pages} {513} (\bibinfo {year} {2018})}\BibitemShut {NoStop}%
\bibitem [{\citenamefont {Young}\ \emph {et~al.}(2018)\citenamefont {Young},
  \citenamefont {Hundt}, \citenamefont {Cole}, \citenamefont {Fineberg},
  \citenamefont {Andrecka}, \citenamefont {Tyler}, \citenamefont {Olerinyova},
  \citenamefont {Ansari}, \citenamefont {Marklund}, \citenamefont {Collier},
  \citenamefont {Chandler}, \citenamefont {Tkachenko}, \citenamefont {Allen},
  \citenamefont {Crispin}, \citenamefont {Billington}, \citenamefont {Takagi},
  \citenamefont {Sellers}, \citenamefont {Eichmann}, \citenamefont {Selenko},
  \citenamefont {Frey}, \citenamefont {Riek}, \citenamefont {Galpin},
  \citenamefont {Struwe}, \citenamefont {Benesch},\ and\ \citenamefont
  {Kukura}}]{Young2018}%
  \BibitemOpen
  \bibfield  {author} {\bibinfo {author} {\bibfnamefont {G.}~\bibnamefont
  {Young}}, \bibinfo {author} {\bibfnamefont {N.}~\bibnamefont {Hundt}},
  \bibinfo {author} {\bibfnamefont {D.}~\bibnamefont {Cole}}, \bibinfo {author}
  {\bibfnamefont {A.}~\bibnamefont {Fineberg}}, \bibinfo {author}
  {\bibfnamefont {J.}~\bibnamefont {Andrecka}}, \bibinfo {author}
  {\bibfnamefont {A.}~\bibnamefont {Tyler}}, \bibinfo {author} {\bibfnamefont
  {A.}~\bibnamefont {Olerinyova}}, \bibinfo {author} {\bibfnamefont
  {A.}~\bibnamefont {Ansari}}, \bibinfo {author} {\bibfnamefont {E.~G.}\
  \bibnamefont {Marklund}}, \bibinfo {author} {\bibfnamefont {M.~P.}\
  \bibnamefont {Collier}}, \bibinfo {author} {\bibfnamefont {S.~A.}\
  \bibnamefont {Chandler}}, \bibinfo {author} {\bibfnamefont {O.}~\bibnamefont
  {Tkachenko}}, \bibinfo {author} {\bibfnamefont {J.}~\bibnamefont {Allen}},
  \bibinfo {author} {\bibfnamefont {M.}~\bibnamefont {Crispin}}, \bibinfo
  {author} {\bibfnamefont {N.}~\bibnamefont {Billington}}, \bibinfo {author}
  {\bibfnamefont {Y.}~\bibnamefont {Takagi}}, \bibinfo {author} {\bibfnamefont
  {J.~R.}\ \bibnamefont {Sellers}}, \bibinfo {author} {\bibfnamefont
  {C.}~\bibnamefont {Eichmann}}, \bibinfo {author} {\bibfnamefont
  {P.}~\bibnamefont {Selenko}}, \bibinfo {author} {\bibfnamefont
  {L.}~\bibnamefont {Frey}}, \bibinfo {author} {\bibfnamefont {R.}~\bibnamefont
  {Riek}}, \bibinfo {author} {\bibfnamefont {M.~R.}\ \bibnamefont {Galpin}},
  \bibinfo {author} {\bibfnamefont {W.~B.}\ \bibnamefont {Struwe}}, \bibinfo
  {author} {\bibfnamefont {J.~L.~P.}\ \bibnamefont {Benesch}},\ and\ \bibinfo
  {author} {\bibfnamefont {P.}~\bibnamefont {Kukura}},\ }\bibfield  {title}
  {\bibinfo {title} {{Quantitative Mass Imaging of Single Biological
  Macromolecules.}},\ }\href {https://doi.org/10.1126/science.aar5839}
  {\bibfield  {journal} {\bibinfo  {journal} {Science}\ }\textbf {\bibinfo
  {volume} {360}},\ \bibinfo {pages} {423} (\bibinfo {year}
  {2018})}\BibitemShut {NoStop}%
\bibitem [{\citenamefont {Zijlstra}\ \emph {et~al.}(2012)\citenamefont
  {Zijlstra}, \citenamefont {Paulo},\ and\ \citenamefont
  {Orrit}}]{Zijlstra_2012}%
  \BibitemOpen
  \bibfield  {author} {\bibinfo {author} {\bibfnamefont {P.}~\bibnamefont
  {Zijlstra}}, \bibinfo {author} {\bibfnamefont {P.~M.~R.}\ \bibnamefont
  {Paulo}},\ and\ \bibinfo {author} {\bibfnamefont {M.}~\bibnamefont {Orrit}},\
  }\bibfield  {title} {\bibinfo {title} {Optical detection of single
  non-absorbing molecules using the surface plasmon resonance of a gold
  nanorod},\ }\href {https://doi.org/10.1038/nnano.2012.51} {\bibfield
  {journal} {\bibinfo  {journal} {Nat. Nanotechnol.}\ }\textbf {\bibinfo
  {volume} {7}},\ \bibinfo {pages} {379} (\bibinfo {year} {2012})}\BibitemShut
  {NoStop}%
\bibitem [{\citenamefont {Ament}\ \emph {et~al.}(2012)\citenamefont {Ament},
  \citenamefont {Prasad}, \citenamefont {Henkel}, \citenamefont {Schmachtel},\
  and\ \citenamefont {S\"onnichsen}}]{Ament2012}%
  \BibitemOpen
  \bibfield  {author} {\bibinfo {author} {\bibfnamefont {I.}~\bibnamefont
  {Ament}}, \bibinfo {author} {\bibfnamefont {J.}~\bibnamefont {Prasad}},
  \bibinfo {author} {\bibfnamefont {A.}~\bibnamefont {Henkel}}, \bibinfo
  {author} {\bibfnamefont {S.}~\bibnamefont {Schmachtel}},\ and\ \bibinfo
  {author} {\bibfnamefont {C.}~\bibnamefont {S\"onnichsen}},\ }\bibfield
  {title} {\bibinfo {title} {Single unlabeled protein detection on individual
  plasmonic nanoparticles},\ }\href {https://doi.org/10.1021/nl204496g}
  {\bibfield  {journal} {\bibinfo  {journal} {Nano Letters}\ }\textbf {\bibinfo
  {volume} {12}},\ \bibinfo {pages} {1092} (\bibinfo {year}
  {2012})}\BibitemShut {NoStop}%
\bibitem [{\citenamefont {Rosman}\ \emph {et~al.}(2013)\citenamefont {Rosman},
  \citenamefont {Prasad}, \citenamefont {Neiser}, \citenamefont {Henkel},
  \citenamefont {Edgar},\ and\ \citenamefont {S\"onnichsen}}]{Rosman2013}%
  \BibitemOpen
  \bibfield  {author} {\bibinfo {author} {\bibfnamefont {C.}~\bibnamefont
  {Rosman}}, \bibinfo {author} {\bibfnamefont {J.}~\bibnamefont {Prasad}},
  \bibinfo {author} {\bibfnamefont {A.}~\bibnamefont {Neiser}}, \bibinfo
  {author} {\bibfnamefont {A.}~\bibnamefont {Henkel}}, \bibinfo {author}
  {\bibfnamefont {J.}~\bibnamefont {Edgar}},\ and\ \bibinfo {author}
  {\bibfnamefont {C.}~\bibnamefont {S\"onnichsen}},\ }\bibfield  {title}
  {\bibinfo {title} {Multiplexed plasmon sensor for rapid label-free analyte
  detection},\ }\href {https://doi.org/10.1021/nl401354f} {\bibfield  {journal}
  {\bibinfo  {journal} {Nano Letters}\ }\textbf {\bibinfo {volume} {13}},\
  \bibinfo {pages} {3243} (\bibinfo {year} {2013})}\BibitemShut {NoStop}%
\bibitem [{\citenamefont {Beuwer}\ \emph {et~al.}(2015)\citenamefont {Beuwer},
  \citenamefont {Prins},\ and\ \citenamefont {Zijlstra}}]{Beuwer2015}%
  \BibitemOpen
  \bibfield  {author} {\bibinfo {author} {\bibfnamefont {M.~A.}\ \bibnamefont
  {Beuwer}}, \bibinfo {author} {\bibfnamefont {M.~W.~J.}\ \bibnamefont
  {Prins}},\ and\ \bibinfo {author} {\bibfnamefont {P.}~\bibnamefont
  {Zijlstra}},\ }\bibfield  {title} {\bibinfo {title} {Stochastic protein
  interactions monitored by hundreds of single-molecule plasmonic biosensors},\
  }\href {https://doi.org/10.1021/acs.nanolett.5b00872} {\bibfield  {journal}
  {\bibinfo  {journal} {Nano Letters}\ }\textbf {\bibinfo {volume} {15}},\
  \bibinfo {pages} {3507} (\bibinfo {year} {2015})}\BibitemShut {NoStop}%
\bibitem [{\citenamefont {Ye}\ \emph {et~al.}(2018)\citenamefont {Ye},
  \citenamefont {G\"otz}, \citenamefont {Celiksoy}, \citenamefont {T\"uting},
  \citenamefont {Ratzke}, \citenamefont {Prasad}, \citenamefont {Ricken},
  \citenamefont {Wegner}, \citenamefont {Ahijado-Guzm\'an}, \citenamefont
  {Hugel},\ and\ \citenamefont {S\"onnichsen}}]{Ye2018}%
  \BibitemOpen
  \bibfield  {author} {\bibinfo {author} {\bibfnamefont {W.}~\bibnamefont
  {Ye}}, \bibinfo {author} {\bibfnamefont {M.}~\bibnamefont {G\"otz}}, \bibinfo
  {author} {\bibfnamefont {S.}~\bibnamefont {Celiksoy}}, \bibinfo {author}
  {\bibfnamefont {L.}~\bibnamefont {T\"uting}}, \bibinfo {author}
  {\bibfnamefont {C.}~\bibnamefont {Ratzke}}, \bibinfo {author} {\bibfnamefont
  {J.}~\bibnamefont {Prasad}}, \bibinfo {author} {\bibfnamefont
  {J.}~\bibnamefont {Ricken}}, \bibinfo {author} {\bibfnamefont {S.~V.}\
  \bibnamefont {Wegner}}, \bibinfo {author} {\bibfnamefont {R.}~\bibnamefont
  {Ahijado-Guzm\'an}}, \bibinfo {author} {\bibfnamefont {T.}~\bibnamefont
  {Hugel}},\ and\ \bibinfo {author} {\bibfnamefont {C.}~\bibnamefont
  {S\"onnichsen}},\ }\bibfield  {title} {\bibinfo {title} {Conformational
  dynamics of a single protein monitored for 24 h at video rate},\ }\href
  {https://doi.org/10.1021/acs.nanolett.8b03342} {\bibfield  {journal}
  {\bibinfo  {journal} {Nano Letters}\ }\textbf {\bibinfo {volume} {18}},\
  \bibinfo {pages} {6633} (\bibinfo {year} {2018})}\BibitemShut {NoStop}%
\bibitem [{\citenamefont {Vollmer}\ \emph {et~al.}(2008)\citenamefont
  {Vollmer}, \citenamefont {Arnold},\ and\ \citenamefont {Keng}}]{Vollmer2008}%
  \BibitemOpen
  \bibfield  {author} {\bibinfo {author} {\bibfnamefont {F.}~\bibnamefont
  {Vollmer}}, \bibinfo {author} {\bibfnamefont {S.}~\bibnamefont {Arnold}},\
  and\ \bibinfo {author} {\bibfnamefont {D.}~\bibnamefont {Keng}},\ }\bibfield
  {title} {\bibinfo {title} {{Single Virus Detection from the Reactive Shift of
  a Whispering-Gallery Mode}},\ }\href
  {https://doi.org/10.1073/pnas.0808988106} {\bibfield  {journal} {\bibinfo
  {journal} {Proc. Natl. Acad. Sci. U. S. A.}\ }\textbf {\bibinfo {volume}
  {105}},\ \bibinfo {pages} {20701} (\bibinfo {year} {2008})}\BibitemShut
  {NoStop}%
\bibitem [{\citenamefont {Zhu}\ \emph {et~al.}(2010)\citenamefont {Zhu},
  \citenamefont {Ozdemir}, \citenamefont {Xiao}, \citenamefont {Li},
  \citenamefont {He}, \citenamefont {Chen},\ and\ \citenamefont
  {Yang}}]{Zhu2009}%
  \BibitemOpen
  \bibfield  {author} {\bibinfo {author} {\bibfnamefont {J.}~\bibnamefont
  {Zhu}}, \bibinfo {author} {\bibfnamefont {S.~K.}\ \bibnamefont {Ozdemir}},
  \bibinfo {author} {\bibfnamefont {Y.-F.}\ \bibnamefont {Xiao}}, \bibinfo
  {author} {\bibfnamefont {L.}~\bibnamefont {Li}}, \bibinfo {author}
  {\bibfnamefont {L.}~\bibnamefont {He}}, \bibinfo {author} {\bibfnamefont
  {D.-R.}\ \bibnamefont {Chen}},\ and\ \bibinfo {author} {\bibfnamefont
  {L.}~\bibnamefont {Yang}},\ }\bibfield  {title} {\bibinfo {title} {On-chip
  single nanoparticle detection and sizing by mode splitting in an ultrahigh-q
  microresonator},\ }\href {https://doi.org/10.1038/nphoton.2009.237}
  {\bibfield  {journal} {\bibinfo  {journal} {Nat. Photonics}\ }\textbf
  {\bibinfo {volume} {4}},\ \bibinfo {pages} {46} (\bibinfo {year}
  {2010})}\BibitemShut {NoStop}%
\bibitem [{\citenamefont {He}\ \emph {et~al.}(2011)\citenamefont {He},
  \citenamefont {Özdemir}, \citenamefont {Zhu}, \citenamefont {Kim},\ and\
  \citenamefont {Yang}}]{He2011}%
  \BibitemOpen
  \bibfield  {author} {\bibinfo {author} {\bibfnamefont {L.}~\bibnamefont
  {He}}, \bibinfo {author} {\bibfnamefont {S.~K.}\ \bibnamefont {Özdemir}},
  \bibinfo {author} {\bibfnamefont {J.}~\bibnamefont {Zhu}}, \bibinfo {author}
  {\bibfnamefont {W.}~\bibnamefont {Kim}},\ and\ \bibinfo {author}
  {\bibfnamefont {L.}~\bibnamefont {Yang}},\ }\bibfield  {title} {\bibinfo
  {title} {{Detecting Single Viruses and Nanoparticles using Whispering Gallery
  Microlasers}},\ }\href {https://doi.org/10.1038/NNANO.2011.99} {\bibfield
  {journal} {\bibinfo  {journal} {Nat. Nanotechnol.}\ }\textbf {\bibinfo
  {volume} {6}},\ \bibinfo {pages} {428} (\bibinfo {year} {2011})}\BibitemShut
  {NoStop}%
\bibitem [{\citenamefont {Dantham}\ \emph {et~al.}(2013)\citenamefont
  {Dantham}, \citenamefont {Holler}, \citenamefont {Barbre}, \citenamefont
  {Keng}, \citenamefont {Kolchenko},\ and\ \citenamefont
  {Arnold}}]{Dantham2013}%
  \BibitemOpen
  \bibfield  {author} {\bibinfo {author} {\bibfnamefont {V.~R.}\ \bibnamefont
  {Dantham}}, \bibinfo {author} {\bibfnamefont {S.}~\bibnamefont {Holler}},
  \bibinfo {author} {\bibfnamefont {C.}~\bibnamefont {Barbre}}, \bibinfo
  {author} {\bibfnamefont {D.}~\bibnamefont {Keng}}, \bibinfo {author}
  {\bibfnamefont {V.}~\bibnamefont {Kolchenko}},\ and\ \bibinfo {author}
  {\bibfnamefont {S.}~\bibnamefont {Arnold}},\ }\bibfield  {title} {\bibinfo
  {title} {Label-free detection of single protein using a
  nanoplasmonic-photonic hybrid microcavity},\ }\href
  {https://doi.org/10.1021/nl401633y} {\bibfield  {journal} {\bibinfo
  {journal} {Nano Letters}\ }\textbf {\bibinfo {volume} {13}},\ \bibinfo
  {pages} {3347} (\bibinfo {year} {2013})}\BibitemShut {NoStop}%
\bibitem [{\citenamefont {Li}\ \emph {et~al.}(2014)\citenamefont {Li},
  \citenamefont {Clements}, \citenamefont {Yu}, \citenamefont {Shi},
  \citenamefont {Gong},\ and\ \citenamefont {Xiao}}]{Bei-Bei2014}%
  \BibitemOpen
  \bibfield  {author} {\bibinfo {author} {\bibfnamefont {B.-B.}\ \bibnamefont
  {Li}}, \bibinfo {author} {\bibfnamefont {W.~R.}\ \bibnamefont {Clements}},
  \bibinfo {author} {\bibfnamefont {X.-C.}\ \bibnamefont {Yu}}, \bibinfo
  {author} {\bibfnamefont {K.}~\bibnamefont {Shi}}, \bibinfo {author}
  {\bibfnamefont {Q.}~\bibnamefont {Gong}},\ and\ \bibinfo {author}
  {\bibfnamefont {Y.-F.}\ \bibnamefont {Xiao}},\ }\bibfield  {title} {\bibinfo
  {title} {Single nanoparticle detection using split-mode microcavity raman
  lasers},\ }\href {https://doi.org/10.1073/pnas.1408453111} {\bibfield
  {journal} {\bibinfo  {journal} {Proc. Natl. Acad. Sci. U. S. A.}\ }\textbf
  {\bibinfo {volume} {111}},\ \bibinfo {pages} {14657} (\bibinfo {year}
  {2014})}\BibitemShut {NoStop}%
\bibitem [{\citenamefont {Baaske}\ \emph {et~al.}(2014)\citenamefont {Baaske},
  \citenamefont {Foreman},\ and\ \citenamefont {Vollmer}}]{Baaske2014}%
  \BibitemOpen
  \bibfield  {author} {\bibinfo {author} {\bibfnamefont {M.~D.}\ \bibnamefont
  {Baaske}}, \bibinfo {author} {\bibfnamefont {M.~R.}\ \bibnamefont
  {Foreman}},\ and\ \bibinfo {author} {\bibfnamefont {F.}~\bibnamefont
  {Vollmer}},\ }\bibfield  {title} {\bibinfo {title} {Single-molecule nucleic
  acid interactions monitored on a label-free microcavity biosensor platform},\
  }\href {https://doi.org/10.1038/nnano.2014.180} {\bibfield  {journal}
  {\bibinfo  {journal} {Nat. Nanotechnol.}\ }\textbf {\bibinfo {volume} {9}},\
  \bibinfo {pages} {933} (\bibinfo {year} {2014})}\BibitemShut {NoStop}%
\bibitem [{\citenamefont {Kim}\ \emph {et~al.}(2016)\citenamefont {Kim},
  \citenamefont {Baaske},\ and\ \citenamefont {Vollmer}}]{Kim2016}%
  \BibitemOpen
  \bibfield  {author} {\bibinfo {author} {\bibfnamefont {E.}~\bibnamefont
  {Kim}}, \bibinfo {author} {\bibfnamefont {M.~D.}\ \bibnamefont {Baaske}},\
  and\ \bibinfo {author} {\bibfnamefont {F.}~\bibnamefont {Vollmer}},\
  }\bibfield  {title} {\bibinfo {title} {In situ observation of single-molecule
  surface reactions from low to high affinities},\ }\href
  {https://doi.org/10.1002/adma.201603153} {\bibfield  {journal} {\bibinfo
  {journal} {Advanced Materials}\ }\textbf {\bibinfo {volume} {28}},\ \bibinfo
  {pages} {9941} (\bibinfo {year} {2016})}\BibitemShut {NoStop}%
\bibitem [{\citenamefont {Baaske}\ and\ \citenamefont
  {Vollmer}(2016)}]{Baaske2016}%
  \BibitemOpen
  \bibfield  {author} {\bibinfo {author} {\bibfnamefont {M.~D.}\ \bibnamefont
  {Baaske}}\ and\ \bibinfo {author} {\bibfnamefont {F.}~\bibnamefont
  {Vollmer}},\ }\bibfield  {title} {\bibinfo {title} {Optical observation of
  single atomic ions interacting with plasmonic nanorods in aqueous solution},\
  }\href {https://doi.org/10.1038/nphoton.2016.177} {\bibfield  {journal}
  {\bibinfo  {journal} {Nat. Photonics}\ }\textbf {\bibinfo {volume} {10}},\
  \bibinfo {pages} {733} (\bibinfo {year} {2016})}\BibitemShut {NoStop}%
\bibitem [{\citenamefont {Chen}\ \emph {et~al.}(2017)\citenamefont {Chen},
  \citenamefont {Özdemir}, \citenamefont {Zhao}, \citenamefont {Wiersig},\
  and\ \citenamefont {Yang}}]{Chen2017}%
  \BibitemOpen
  \bibfield  {author} {\bibinfo {author} {\bibfnamefont {W.}~\bibnamefont
  {Chen}}, \bibinfo {author} {\bibfnamefont {S.~K.}\ \bibnamefont {Özdemir}},
  \bibinfo {author} {\bibfnamefont {G.}~\bibnamefont {Zhao}}, \bibinfo {author}
  {\bibfnamefont {J.}~\bibnamefont {Wiersig}},\ and\ \bibinfo {author}
  {\bibfnamefont {L.}~\bibnamefont {Yang}},\ }\bibfield  {title} {\bibinfo
  {title} {{Exceptional Points Enhance Sensing in an Optical Microcavity}},\
  }\href {https://doi.org/10.1038/nature232} {\bibfield  {journal} {\bibinfo
  {journal} {Nature}\ }\textbf {\bibinfo {volume} {548}},\ \bibinfo {pages}
  {192} (\bibinfo {year} {2017})}\BibitemShut {NoStop}%
\bibitem [{\citenamefont {Kim}\ \emph {et~al.}(2017)\citenamefont {Kim},
  \citenamefont {Baaske}, \citenamefont {Schuldes}, \citenamefont {Wilsch},\
  and\ \citenamefont {Vollmer}}]{Kim2017}%
  \BibitemOpen
  \bibfield  {author} {\bibinfo {author} {\bibfnamefont {E.}~\bibnamefont
  {Kim}}, \bibinfo {author} {\bibfnamefont {M.~D.}\ \bibnamefont {Baaske}},
  \bibinfo {author} {\bibfnamefont {I.}~\bibnamefont {Schuldes}}, \bibinfo
  {author} {\bibfnamefont {P.~S.}\ \bibnamefont {Wilsch}},\ and\ \bibinfo
  {author} {\bibfnamefont {F.}~\bibnamefont {Vollmer}},\ }\bibfield  {title}
  {\bibinfo {title} {Label-free optical detection of single enzyme-reactant
  reactions and associated conformational changes},\ }\bibfield  {journal}
  {\bibinfo  {journal} {Science Advances}\ }\textbf {\bibinfo {volume} {3}},\
  \href {https://doi.org/10.1126/sciadv.1603044} {10.1126/sciadv.1603044}
  (\bibinfo {year} {2017})\BibitemShut {NoStop}%
\bibitem [{\citenamefont {Ignatovich}\ and\ \citenamefont
  {Novotny}(2006)}]{Ignatovich2006a}%
  \BibitemOpen
  \bibfield  {author} {\bibinfo {author} {\bibfnamefont {F.~V.}\ \bibnamefont
  {Ignatovich}}\ and\ \bibinfo {author} {\bibfnamefont {L.}~\bibnamefont
  {Novotny}},\ }\bibfield  {title} {\bibinfo {title} {{Real-Time and
  Background-Free Detection of Nanoscale Particles}},\ }\href
  {https://doi.org/10.1103/PhysRevLett.96.013901} {\bibfield  {journal}
  {\bibinfo  {journal} {Phys. Rev. Lett.}\ }\textbf {\bibinfo {volume} {96}},\
  \bibinfo {pages} {1} (\bibinfo {year} {2006})}\BibitemShut {NoStop}%
\bibitem [{\citenamefont {Krishnan}\ \emph {et~al.}(2010)\citenamefont
  {Krishnan}, \citenamefont {Mojarad}, \citenamefont {Kukura},\ and\
  \citenamefont {Sandoghdar}}]{Krishnan2010}%
  \BibitemOpen
  \bibfield  {author} {\bibinfo {author} {\bibfnamefont {M.}~\bibnamefont
  {Krishnan}}, \bibinfo {author} {\bibfnamefont {N.}~\bibnamefont {Mojarad}},
  \bibinfo {author} {\bibfnamefont {P.}~\bibnamefont {Kukura}},\ and\ \bibinfo
  {author} {\bibfnamefont {V.}~\bibnamefont {Sandoghdar}},\ }\bibfield  {title}
  {\bibinfo {title} {{Geometry-Induced Electrostatic Trapping of Nanometric
  Objects in a Fluid}},\ }\href@noop {} {\bibfield  {journal} {\bibinfo
  {journal} {Nature}\ }\textbf {\bibinfo {volume} {467}},\ \bibinfo {pages}
  {692} (\bibinfo {year} {2010})}\BibitemShut {NoStop}%
\bibitem [{\citenamefont {Faez}\ \emph {et~al.}(2015)\citenamefont {Faez},
  \citenamefont {Lahini}, \citenamefont {Weidlich}, \citenamefont {Garmann},
  \citenamefont {Wondraczek}, \citenamefont {Zeisberger}, \citenamefont
  {Schmidt}, \citenamefont {Orrit},\ and\ \citenamefont
  {Manoharan}}]{Faez2015}%
  \BibitemOpen
  \bibfield  {author} {\bibinfo {author} {\bibfnamefont {S.}~\bibnamefont
  {Faez}}, \bibinfo {author} {\bibfnamefont {Y.}~\bibnamefont {Lahini}},
  \bibinfo {author} {\bibfnamefont {S.}~\bibnamefont {Weidlich}}, \bibinfo
  {author} {\bibfnamefont {R.~F.}\ \bibnamefont {Garmann}}, \bibinfo {author}
  {\bibfnamefont {K.}~\bibnamefont {Wondraczek}}, \bibinfo {author}
  {\bibfnamefont {M.}~\bibnamefont {Zeisberger}}, \bibinfo {author}
  {\bibfnamefont {M.~A.}\ \bibnamefont {Schmidt}}, \bibinfo {author}
  {\bibfnamefont {M.}~\bibnamefont {Orrit}},\ and\ \bibinfo {author}
  {\bibfnamefont {V.~N.}\ \bibnamefont {Manoharan}},\ }\bibfield  {title}
  {\bibinfo {title} {Fast, label-free tracking of single viruses and weakly
  scattering nanoparticles in a nanofluidic optical fiber},\ }\href
  {https://doi.org/10.1021/acsnano.5b05646} {\bibfield  {journal} {\bibinfo
  {journal} {ACS Nano}\ }\textbf {\bibinfo {volume} {9}},\ \bibinfo {pages}
  {12349} (\bibinfo {year} {2015})}\BibitemShut {NoStop}%
\bibitem [{\citenamefont {Squires}\ \emph {et~al.}(2019)\citenamefont
  {Squires}, \citenamefont {Lavania}, \citenamefont {Dahlberg},\ and\
  \citenamefont {Moerner}}]{Squire2019}%
  \BibitemOpen
  \bibfield  {author} {\bibinfo {author} {\bibfnamefont {A.~H.}\ \bibnamefont
  {Squires}}, \bibinfo {author} {\bibfnamefont {A.~A.}\ \bibnamefont
  {Lavania}}, \bibinfo {author} {\bibfnamefont {P.~D.}\ \bibnamefont
  {Dahlberg}},\ and\ \bibinfo {author} {\bibfnamefont {W.~E.}\ \bibnamefont
  {Moerner}},\ }\bibfield  {title} {\bibinfo {title} {Interferometric
  scattering enables fluorescence-free electrokinetic trapping of single
  nanoparticles in free solution},\ }\href
  {https://doi.org/10.1021/acs.nanolett.9b01514} {\bibfield  {journal}
  {\bibinfo  {journal} {Nano Letters}\ }\textbf {\bibinfo {volume} {19}},\
  \bibinfo {pages} {4112} (\bibinfo {year} {2019})}\BibitemShut {NoStop}%
\bibitem [{\citenamefont {Wang}\ \emph {et~al.}(2005)\citenamefont {Wang},
  \citenamefont {Huff}, \citenamefont {Zweifel}, \citenamefont {He},
  \citenamefont {Low}, \citenamefont {Wei},\ and\ \citenamefont
  {Cheng}}]{Wang2005}%
  \BibitemOpen
  \bibfield  {author} {\bibinfo {author} {\bibfnamefont {H.}~\bibnamefont
  {Wang}}, \bibinfo {author} {\bibfnamefont {T.~B.}\ \bibnamefont {Huff}},
  \bibinfo {author} {\bibfnamefont {D.~A.}\ \bibnamefont {Zweifel}}, \bibinfo
  {author} {\bibfnamefont {W.}~\bibnamefont {He}}, \bibinfo {author}
  {\bibfnamefont {P.~S.}\ \bibnamefont {Low}}, \bibinfo {author} {\bibfnamefont
  {A.}~\bibnamefont {Wei}},\ and\ \bibinfo {author} {\bibfnamefont {J.-X.}\
  \bibnamefont {Cheng}},\ }\bibfield  {title} {\bibinfo {title} {{In vitro and
  in vivo two-photon luminescence imaging of single gold nanorods}},\
  }\href@noop {} {\bibfield  {journal} {\bibinfo  {journal} {Proceedings of the
  National Academy of Sciences of the United States of America}\ }\textbf
  {\bibinfo {volume} {102}},\ \bibinfo {pages} {15752} (\bibinfo {year}
  {2005})}\BibitemShut {NoStop}%
\bibitem [{\citenamefont {Huang}\ \emph {et~al.}(2006)\citenamefont {Huang},
  \citenamefont {El-Sayed}, \citenamefont {Qian},\ and\ \citenamefont
  {El-Sayed}}]{Huang2006}%
  \BibitemOpen
  \bibfield  {author} {\bibinfo {author} {\bibfnamefont {X.}~\bibnamefont
  {Huang}}, \bibinfo {author} {\bibfnamefont {I.~H.}\ \bibnamefont {El-Sayed}},
  \bibinfo {author} {\bibfnamefont {W.}~\bibnamefont {Qian}},\ and\ \bibinfo
  {author} {\bibfnamefont {M.~A.}\ \bibnamefont {El-Sayed}},\ }\bibfield
  {title} {\bibinfo {title} {Cancer cell imaging and photothermal therapy in
  the near-infrared region by using gold nanorods},\ }\href
  {https://doi.org/10.1021/ja057254a} {\bibfield  {journal} {\bibinfo
  {journal} {Journal of the American Chemical Society}\ }\textbf {\bibinfo
  {volume} {128}},\ \bibinfo {pages} {2115} (\bibinfo {year}
  {2006})}\BibitemShut {NoStop}%
\bibitem [{\citenamefont {Durr}\ \emph {et~al.}(2007)\citenamefont {Durr},
  \citenamefont {Larson}, \citenamefont {Smith}, \citenamefont {Korgel},
  \citenamefont {Sokolov},\ and\ \citenamefont {Ben-Yakar}}]{Durr2007}%
  \BibitemOpen
  \bibfield  {author} {\bibinfo {author} {\bibfnamefont {N.~J.}\ \bibnamefont
  {Durr}}, \bibinfo {author} {\bibfnamefont {T.}~\bibnamefont {Larson}},
  \bibinfo {author} {\bibfnamefont {D.~K.}\ \bibnamefont {Smith}}, \bibinfo
  {author} {\bibfnamefont {B.~A.}\ \bibnamefont {Korgel}}, \bibinfo {author}
  {\bibfnamefont {K.}~\bibnamefont {Sokolov}},\ and\ \bibinfo {author}
  {\bibfnamefont {A.}~\bibnamefont {Ben-Yakar}},\ }\bibfield  {title} {\bibinfo
  {title} {Two-photon luminescence imaging of cancer cells using molecularly
  targeted gold nanorods},\ }\href {https://doi.org/10.1021/nl062962v}
  {\bibfield  {journal} {\bibinfo  {journal} {Nano Letters}\ }\textbf {\bibinfo
  {volume} {7}},\ \bibinfo {pages} {941} (\bibinfo {year} {2007})}\BibitemShut
  {NoStop}%
\bibitem [{\citenamefont {Wulf}\ \emph {et~al.}(2016)\citenamefont {Wulf},
  \citenamefont {Knoch}, \citenamefont {Speck},\ and\ \citenamefont
  {S\"onnichsen}}]{Wulf2016}%
  \BibitemOpen
  \bibfield  {author} {\bibinfo {author} {\bibfnamefont {V.}~\bibnamefont
  {Wulf}}, \bibinfo {author} {\bibfnamefont {F.}~\bibnamefont {Knoch}},
  \bibinfo {author} {\bibfnamefont {T.}~\bibnamefont {Speck}},\ and\ \bibinfo
  {author} {\bibfnamefont {C.}~\bibnamefont {S\"onnichsen}},\ }\bibfield
  {title} {\bibinfo {title} {Gold nanorods as plasmonic sensors for particle
  diffusion},\ }\href {https://doi.org/10.1021/acs.jpclett.6b02165} {\bibfield
  {journal} {\bibinfo  {journal} {The Journal of Physical Chemistry Letters}\
  }\textbf {\bibinfo {volume} {7}},\ \bibinfo {pages} {4951} (\bibinfo {year}
  {2016})}\BibitemShut {NoStop}%
\bibitem [{\citenamefont {Jollans}\ \emph {et~al.}(2019)\citenamefont
  {Jollans}, \citenamefont {Baaske},\ and\ \citenamefont
  {Orrit}}]{Jollans2019}%
  \BibitemOpen
  \bibfield  {author} {\bibinfo {author} {\bibfnamefont {T.}~\bibnamefont
  {Jollans}}, \bibinfo {author} {\bibfnamefont {M.~D.}\ \bibnamefont
  {Baaske}},\ and\ \bibinfo {author} {\bibfnamefont {M.}~\bibnamefont
  {Orrit}},\ }\bibfield  {title} {\bibinfo {title} {Nonfluorescent optical
  probing of single molecules and nanoparticles},\ }\href
  {https://doi.org/10.1021/acs.jpcc.9b00843} {\bibfield  {journal} {\bibinfo
  {journal} {The Journal of Physical Chemistry C}\ }\textbf {\bibinfo {volume}
  {123}},\ \bibinfo {pages} {14107} (\bibinfo {year} {2019})}\BibitemShut
  {NoStop}%
\bibitem [{\citenamefont {Warisnoicharoen}\ \emph
  {et~al.}(2000{\natexlab{a}})\citenamefont {Warisnoicharoen}, \citenamefont
  {Lansley},\ and\ \citenamefont {Lawrence}}]{Warisnoicharoen2000}%
  \BibitemOpen
  \bibfield  {author} {\bibinfo {author} {\bibfnamefont {W.}~\bibnamefont
  {Warisnoicharoen}}, \bibinfo {author} {\bibfnamefont {A.~B.}\ \bibnamefont
  {Lansley}},\ and\ \bibinfo {author} {\bibfnamefont {M.~J.}\ \bibnamefont
  {Lawrence}},\ }\bibfield  {title} {\bibinfo {title} {Light-scattering
  investigations on dilute nonionic oil-in-water microemulsions},\ }\href
  {https://doi.org/10.1208/ps020212} {\bibfield  {journal} {\bibinfo  {journal}
  {AAPS PharmSci}\ }\textbf {\bibinfo {volume} {2}},\ \bibinfo {pages} {16}
  (\bibinfo {year} {2000}{\natexlab{a}})}\BibitemShut {NoStop}%
\bibitem [{\citenamefont {Warisnoicharoen}\ \emph
  {et~al.}(2000{\natexlab{b}})\citenamefont {Warisnoicharoen}, \citenamefont
  {Lansley},\ and\ \citenamefont {Lawrence}}]{WARISNOICHAROEN2000_2}%
  \BibitemOpen
  \bibfield  {author} {\bibinfo {author} {\bibfnamefont {W.}~\bibnamefont
  {Warisnoicharoen}}, \bibinfo {author} {\bibfnamefont {A.}~\bibnamefont
  {Lansley}},\ and\ \bibinfo {author} {\bibfnamefont {M.}~\bibnamefont
  {Lawrence}},\ }\bibfield  {title} {\bibinfo {title} {Nonionic oil-in-water
  microemulsions: the effect of oil type on phase behaviour},\ }\href
  {https://doi.org/https://doi.org/10.1016/S0378-5173(99)00406-8} {\bibfield
  {journal} {\bibinfo  {journal} {International Journal of Pharmaceutics}\
  }\textbf {\bibinfo {volume} {198}},\ \bibinfo {pages} {7 } (\bibinfo {year}
  {2000}{\natexlab{b}})}\BibitemShut {NoStop}%
\bibitem [{\citenamefont {Malcolmson}\ \emph {et~al.}(2002)\citenamefont
  {Malcolmson}, \citenamefont {Barlow},\ and\ \citenamefont
  {Lawrence}}]{Malcolmson2002}%
  \BibitemOpen
  \bibfield  {author} {\bibinfo {author} {\bibfnamefont {C.}~\bibnamefont
  {Malcolmson}}, \bibinfo {author} {\bibfnamefont {D.~J.}\ \bibnamefont
  {Barlow}},\ and\ \bibinfo {author} {\bibfnamefont {M.~J.}\ \bibnamefont
  {Lawrence}},\ }\bibfield  {title} {\bibinfo {title} {Light-scattering studies
  of testosterone enanthate containing soybean oil/c18:1e10/water oil-in-water
  microemulsions},\ }\href {https://doi.org/10.1002/jps.10221} {\bibfield
  {journal} {\bibinfo  {journal} {Journal of Pharmaceutical Sciences}\ }\textbf
  {\bibinfo {volume} {91}},\ \bibinfo {pages} {2317} (\bibinfo {year}
  {2002})}\BibitemShut {NoStop}%
\bibitem [{\citenamefont {Benz}\ \emph {et~al.}(2016)\citenamefont {Benz},
  \citenamefont {Schmidt}, \citenamefont {Dreismann}, \citenamefont
  {Chikkaraddy}, \citenamefont {Zhang}, \citenamefont {Demetriadou},
  \citenamefont {Carnegie}, \citenamefont {Ohadi}, \citenamefont {de~Nijs},
  \citenamefont {Esteban}, \citenamefont {Aizpurua},\ and\ \citenamefont
  {Baumberg}}]{Benz2016}%
  \BibitemOpen
  \bibfield  {author} {\bibinfo {author} {\bibfnamefont {F.}~\bibnamefont
  {Benz}}, \bibinfo {author} {\bibfnamefont {M.~K.}\ \bibnamefont {Schmidt}},
  \bibinfo {author} {\bibfnamefont {A.}~\bibnamefont {Dreismann}}, \bibinfo
  {author} {\bibfnamefont {R.}~\bibnamefont {Chikkaraddy}}, \bibinfo {author}
  {\bibfnamefont {Y.}~\bibnamefont {Zhang}}, \bibinfo {author} {\bibfnamefont
  {A.}~\bibnamefont {Demetriadou}}, \bibinfo {author} {\bibfnamefont
  {C.}~\bibnamefont {Carnegie}}, \bibinfo {author} {\bibfnamefont
  {H.}~\bibnamefont {Ohadi}}, \bibinfo {author} {\bibfnamefont
  {B.}~\bibnamefont {de~Nijs}}, \bibinfo {author} {\bibfnamefont
  {R.}~\bibnamefont {Esteban}}, \bibinfo {author} {\bibfnamefont
  {J.}~\bibnamefont {Aizpurua}},\ and\ \bibinfo {author} {\bibfnamefont
  {J.~J.}\ \bibnamefont {Baumberg}},\ }\bibfield  {title} {\bibinfo {title}
  {Single-molecule optomechanics in
  {\textquotedblleft}picocavities{\textquotedblright}},\ }\href
  {https://doi.org/10.1126/science.aah5243} {\bibfield  {journal} {\bibinfo
  {journal} {Science}\ }\textbf {\bibinfo {volume} {354}},\ \bibinfo {pages}
  {726} (\bibinfo {year} {2016})}\BibitemShut {NoStop}%
\bibitem [{\citenamefont {Israelachvili}(2011)}]{Israelachvili2011}%
  \BibitemOpen
  \bibfield  {author} {\bibinfo {author} {\bibfnamefont {J.~N.}\ \bibnamefont
  {Israelachvili}},\ }\href@noop {} {\emph {\bibinfo {title} {Intermolecular
  and Surface Forces}}}\ (\bibinfo  {publisher} {Academic Press},\ \bibinfo
  {year} {2011})\BibitemShut {NoStop}%
\bibitem [{\citenamefont {Caldarola}\ \emph {et~al.}(2018)\citenamefont
  {Caldarola}, \citenamefont {Pradhan},\ and\ \citenamefont
  {Orrit}}]{Caldarola2018}%
  \BibitemOpen
  \bibfield  {author} {\bibinfo {author} {\bibfnamefont {M.}~\bibnamefont
  {Caldarola}}, \bibinfo {author} {\bibfnamefont {B.}~\bibnamefont {Pradhan}},\
  and\ \bibinfo {author} {\bibfnamefont {M.}~\bibnamefont {Orrit}},\ }\bibfield
   {title} {\bibinfo {title} {Quantifying fluorescence enhancement for slowly
  diffusing single molecules in plasmonic near fields},\ }\href
  {https://doi.org/10.1063/1.5023171} {\bibfield  {journal} {\bibinfo
  {journal} {The Journal of Chemical Physics}\ }\textbf {\bibinfo {volume}
  {148}},\ \bibinfo {pages} {123334} (\bibinfo {year} {2018})}\BibitemShut
  {NoStop}%
\end{thebibliography}%


\begin{thebibliography}{4}%
\makeatletter
\providecommand \@ifxundefined [1]{%
 \@ifx{#1\undefined}
}%
\providecommand \@ifnum [1]{%
 \ifnum #1\expandafter \@firstoftwo
 \else \expandafter \@secondoftwo
 \fi
}%
\providecommand \@ifx [1]{%
 \ifx #1\expandafter \@firstoftwo
 \else \expandafter \@secondoftwo
 \fi
}%
\providecommand \natexlab [1]{#1}%
\providecommand \enquote  [1]{``#1''}%
\providecommand \bibnamefont  [1]{#1}%
\providecommand \bibfnamefont [1]{#1}%
\providecommand \citenamefont [1]{#1}%
\providecommand \href@noop [0]{\@secondoftwo}%
\providecommand \href [0]{\begingroup \@sanitize@url \@href}%
\providecommand \@href[1]{\@@startlink{#1}\@@href}%
\providecommand \@@href[1]{\endgroup#1\@@endlink}%
\providecommand \@sanitize@url [0]{\catcode `\\12\catcode `\$12\catcode
  `\&12\catcode `\#12\catcode `\^12\catcode `\_12\catcode `\%12\relax}%
\providecommand \@@startlink[1]{}%
\providecommand \@@endlink[0]{}%
\providecommand \url  [0]{\begingroup\@sanitize@url \@url }%
\providecommand \@url [1]{\endgroup\@href {#1}{\urlprefix }}%
\providecommand \urlprefix  [0]{URL }%
\providecommand \Eprint [0]{\href }%
\providecommand \doibase [0]{https://doi.org/}%
\providecommand \selectlanguage [0]{\@gobble}%
\providecommand \bibinfo  [0]{\@secondoftwo}%
\providecommand \bibfield  [0]{\@secondoftwo}%
\providecommand \translation [1]{[#1]}%
\providecommand \BibitemOpen [0]{}%
\providecommand \bibitemStop [0]{}%
\providecommand \bibitemNoStop [0]{.\EOS\space}%
\providecommand \EOS [0]{\spacefactor3000\relax}%
\providecommand \BibitemShut  [1]{\csname bibitem#1\endcsname}%
\let\auto@bib@innerbib\@empty
\bibitem [{\citenamefont {Bohren}\ and\ \citenamefont
  {Huffman}(1998)}]{Bohren1998}%
  \BibitemOpen
  \bibfield  {author} {\bibinfo {author} {\bibfnamefont {C.}~\bibnamefont
  {Bohren}}\ and\ \bibinfo {author} {\bibfnamefont {D.~R.}\ \bibnamefont
  {Huffman}},\ }\href@noop {} {\emph {\bibinfo {title} {Absorption and
  Scattering of Light by Small Particles}}}\ (\bibinfo  {publisher} {Wiley
  Science Paperback Series},\ \bibinfo {year} {1998})\BibitemShut {NoStop}%
\bibitem [{\citenamefont {Baffou}(2017)}]{Baffou2017}%
  \BibitemOpen
  \bibfield  {author} {\bibinfo {author} {\bibfnamefont {G.}~\bibnamefont
  {Baffou}},\ }\href {https://doi.org/10.1017/9781108289801} {\emph {\bibinfo
  {title} {Thermoplasmonics: Heating Metal Nanoparticles Using Light}}}\
  (\bibinfo  {publisher} {Cambridge University Press},\ \bibinfo {year}
  {2017})\BibitemShut {NoStop}%
\bibitem [{\citenamefont {Israelachvili}(2011)}]{Israelachvili2011}%
  \BibitemOpen
  \bibfield  {author} {\bibinfo {author} {\bibfnamefont {J.~N.}\ \bibnamefont
  {Israelachvili}},\ }\href@noop {} {\emph {\bibinfo {title} {Intermolecular
  and Surface Forces}}}\ (\bibinfo  {publisher} {Academic Press},\ \bibinfo
  {year} {2011})\BibitemShut {NoStop}%
\bibitem [{\citenamefont {Baaske}\ and\ \citenamefont
  {Vollmer}(2016)}]{Baaske2016}%
  \BibitemOpen
  \bibfield  {author} {\bibinfo {author} {\bibfnamefont {M.~D.}\ \bibnamefont
  {Baaske}}\ and\ \bibinfo {author} {\bibfnamefont {F.}~\bibnamefont
  {Vollmer}},\ }\bibfield  {title} {\bibinfo {title} {Optical observation of
  single atomic ions interacting with plasmonic nanorods in aqueous solution},\
  }\href {https://doi.org/10.1038/nphoton.2016.177} {\bibfield  {journal}
  {\bibinfo  {journal} {Nat. Photonics}\ }\textbf {\bibinfo {volume} {10}},\
  \bibinfo {pages} {733} (\bibinfo {year} {2016})}\BibitemShut {NoStop}%
\end{thebibliography}%
\section{Data availability}
The data that support the findings of this study are available from the corresponding author upon reasonable request.

\section{Methods}
\subsection{Setup:}
Here we list the components depicted in figure 1A:
\begin{itemize}
    \setlength\itemsep{0em}
    \item[] Objective: Olympus UPLFLN100XOP
    \item[] Tube lens: Olympus Super Wide Tube Lens Unit
    \item[] Laser: Coherent 890
    \item[] APD: A-Cube S500-240 (Laser Components GmbH)
    \item[] Polarizer LPVISC100 (Thorlabs)
    \item[] 10:90 Beamsplitter BSN11 (Thorlabs)
    \item[] Glan-Thompson Polarizer GTH10M-B (Thorlabs)
    \item[] Piezo Translator P-561.3CD (Phyisk Instrumente GmbH \& Co KG) 
\end{itemize}
Traces were digitized with an oscilloscope (WaveSurfer 24MXs-B, Teledyne Lecroy) and streamed to a PC. Traces with a length of $0.1$\,ms were typically recorded with at a rate of $5\cdot10^8$ samples per second. Consecutive traces were obtained at a rate of $20$ traces per second.
\subsection{Slide preparation:}
CTAB-capped gold nanorods were purchased from Nanopartz. GNR stock solutions containing $10$\,mM CTAB were sonicated (10 min./ Branson 2510) and then deposited onto glass slides (Borosilicate glass diameter $25$\,mm thickness No.1, VWR ) via spin-coating (Specialty coating Systems Spin Coater 6700). The CTAB-layer was consequently removed via UV-cleaning ($15$\,min., Jelight Company Inc. UVO-Cleaner) and the slide was rinsed with Milli-Q water.
\subsection{Preparation of Gold Nanoparticles:}
Citrate-capped $5$\,nm diameter GNPs were purchased from Nanopartz and sonicated for 10 minutes before injection.
\subsection{Preparation of the Microemulsion:}
The preparation of the soybean oil/ polyoxyethylene-10-oleyl ether (Brij-O10)/ water emulsion system was performed in accordance to W. Warisnoicharoen et al. \cite{WARISNOICHAROEN2000_2} i.e. by heating the mixture to $(343\sim353)$\,K for $10$ minutes and consequently cooling it down to $298$\,K, all while continuously stirring the solution.
All chemicals were purchased from Sigma-Aldrich.
Microemulsions were stored at room temperature.

\end{document}


\title{Label-free Plasmonic Detection of Untethered Nanometer-sized Brownian Particles - Supplementary Information}
\author{Martin Dieter Baaske}
\author{Peter Sebastian Neu}
\author{Michel Orrit*}
\affiliation{Huygens-Kamerlingh Onnes Laboratory, Leiden University, Postbus 9504, 2300 RA Leiden, The Netherlands \\ *email: orrit@physics.leidenuniv.nl }
\date{\today}

\maketitle

\subsection{S1: Tuning function $T$}
By utilizing the scattering anisotropy associated with the longitudinal plasmon resonance one can tune the experimental signal to noise ratio via adjustment of input and analyzed polarization with respect to the orientation of the gold nanorod's long axis. Moreover the phase relation between scattered and reflected light can be altered via the detuning of excitation wavelength with respect to the nanorod's resonance. All these factors can be combined into one tuning function.
In the following we will describe the derivation of this tuning function $T$.
Here we will use the effective reflected and the scattered field amplitudes: $\hat{r}= \mathbf {E}_R \cdot\mathbf{n}_{A} $ and $\hat{s}=  \mathbf{E}_S\cdot \mathbf{n}_A$, where $\mathbf{E}_R$, $\mathbf{E}_S$, and $\mathbf{n}_A$ denote the reflected, scattered field, and orientation of the polarization analyzer, respectively.
These effective amplitudes are complex and contain the phase shift of the respective fields with respect to the incident field.
Specifically $\hat{r}=r e^{-i\gamma}$, where $r=|\mathbf{E}_R|\cos{\theta_A}$, $\theta_A$ is the angle between the reflected field's polarization and the analyzer's orientation (compare manuscript Fig. 1B) and $\gamma$ includes the optical phase and part of the Gouy phase. In order to take the resonant behavior of the nanorod into account we describe $\hat{s}$ with a Lorentz oscillator model: \begin{equation}
\hat{s}=-s\frac{1}{\Delta + i},
\end{equation}
where $\Delta=(\nu-\nu_0)/\Gamma$ describes the frequency detuning of the driving laser field's  frequency $\nu$ with respect to the NR's resonance frequency $\nu_0$ normalized to the NR's half width at half maximum $\Gamma$ and   
and $s=|\mathbf{E}_S(\Delta{=}0)|\cos{\theta_S}$ denotes the scattered amplitude on resonance and $\theta_S$ is the angle between the NR's long axis and the analyzer's orientation (comp. manuscript Fig. 1B).  
 
Hence we find for the field amplitude $E_{det}$ after the analyzer: 
\begin{equation}
E_{det}=r e^{-i\gamma}-s\frac{1}{\Delta + i}.
\end{equation}
Consequently we find for the detected power: 
\begin{equation}
P_{det}\propto E_{det}E_{det}^* \propto \left( \left(\frac{r}{s}\right)^2 + \frac{1 + 2\frac{r}{s} \left(  \sin{\gamma} - \Delta \cos{\gamma}\right)}{\Delta^2 + 1}\right) s^2 P_0= \hat{N}^2_s s^2P_0, 
\end{equation}
where $(^*)$ denotes the complex conjugate.
Note that this is intentionally formulated as function of the ratio $\frac{r}{s}$, the main tuning parameter apart from $\Delta$ and $\gamma$.
Our signals are perturbations of this intensity as the resonance of the NR's is shifted by $\delta \nu_0$ and our signal strength is thus determined via :
\begin{equation}
S(\delta \nu_0) \propto \frac{\partial P_{det}}{\partial \nu_0}\delta \nu_0 \propto \left(\frac{2\Delta (1+ 2\frac{r}{s}\sin{\gamma}) +2\frac{r}{s}\cos{\gamma} }{(\Delta^2 + 1)^2}\right)\frac{\delta \nu_0}{\Gamma} s^2P_0=\hat{S}\frac{\delta \nu_0}{\Gamma} s^2P_0
\end{equation}
We describe the measurement noise as:
\begin{equation}
N=\sqrt{h\nu P_{det} B } + N_{BG}\propto (\hat{N}_S +x_{BG}) s \sqrt{h \nu P_0 B}
\end{equation}
where the first term describes the shot noise due to the detected intensity  and bandwidth $B$ and the background term  $N_{BG}$ includes all other noise sources, i.e. electronic noise, laser noise and shot noise from light scattered by impurities and we describe it as fraction of the shot noise caused by the scattered power $s \sqrt{h \nu P_0 B}$ by introducing the factor $x_{BG}$. 
Now we can write the signal to noise ratio as:
\begin{equation}
S/N \propto \frac{\hat{S}}{\hat{N}_s + x_{BG}}s\sqrt{\frac{P_0}{h \nu B}} \frac{\delta\nu_0}{\Gamma}= T(\Delta,\frac{r}{s},\gamma,x_{BG})s\sqrt{\frac{P_0}{h \nu B}} \frac{\delta\nu_0}{\Gamma},
\end{equation}
where $T$ denotes the tuning function, which includes all parameters accessible via adjustment of polarizers and the lasers wavelength as well as the Gouy phase. 
Examples for this tuning function are depicted in figure S1.
\begin{figure}[htbp]
    \centering
    \includegraphics{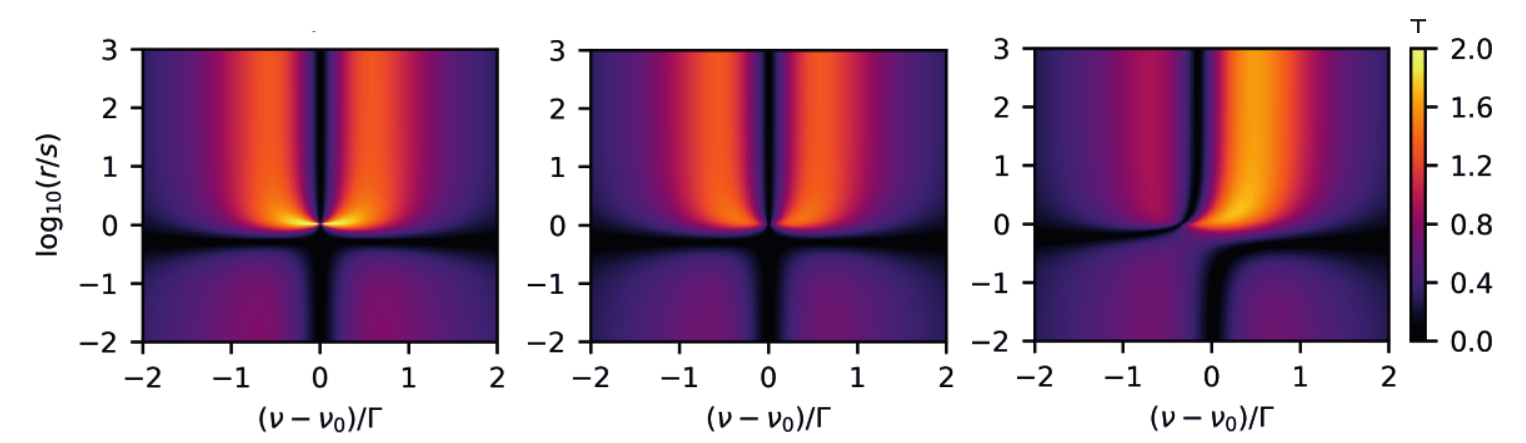}
    \caption{The tuning function $T$ plotted versus frequency detuning and $r/s$-ratio. The left panel shows the confocal  shot-noise-limited case. The center and right panels depict $T$ in presence of background noise ($x_{BG}$) for two values of the Gouy phase $\gamma$.}
    \label{fig:S2}
\end{figure}
Examples for $T$ describing the confocal system ($\gamma=-\pi/2$) are shown in Fig.\,S1. Here, the left graph depicts the ideal shot-noise-limited case which exhibits a singularity at $r/s = 1$ and $\nu=\nu_0$. This describes total destructive interference of scattered and reflected fields. The signal-to-noise ratio becomes infinite as every detected photon originates from a perturbation of the GNR's resonance. The middle panel of Fig.\,S1 shows a more realistic example where the singularity disappears in the presence of background noise ($x_{BG}=0.1$) and the highest $S/N$ values are obtained at the LSPR's flank and $r\geq s$. The right-hand panel shows that even higher $T$-values can be obtained if the Gouy phase deviates slightly from confocal alignment (here $\gamma = -3/5\pi$) and lifts the degeneracy between the short and long wavelength flanks.

\subsection{S2: Nanorod Heating}
Another limitation that has to be considered is heating of the nanorods by our probe beam. Incident powers measured in front of the objective's back focal plane were in the range of $5-20$\,$\mu$W. In order to estimate the absorption cross section $\sigma_{abs}$ of our GNRs we assume prolate ellipsoids with the similar LSPR wavelength and similar volume as our GNRs ($90\times40$\,nm) i.e. an aspect ratio of $3.5$ and a diameter of $38$\,nm. Using the electrostatics approximation\cite{Bohren1998} we obtain $\sigma_{abs}$($\sigma_{scat}$)$=5.4(11.8)\cdot10^{-2}$\,$\mu$m$^2$ on resonance. The FWHM of our setup's point spread function is $0.27$\,$\mu$m. Thus, approximately $5/8$ of the incident power can be absorbed by a single rod. The corresponding increase of the GNRs' temperature would fall into the range of $\Delta T\approx (10-40)$\,K taking into account the GNR's shape and the thermal conductivities of the glass slide and water\cite{Baffou2017}. These $\Delta T$ values, however, reflect the worst-case scenario. If we consider the objective's transmission loss and the off resonance excitation with a polarization that does not match the GNR's orientation we find $\Delta T\lessapprox 8$\,K.

\subsection{S3: Detection of Microemulsion Micelles}
Here we provide additional datasets of our measurements on microemulsion nanodroplets. Specifically, we compare the stretched-exponential fit parameters determined from $G_{100}$\ correlation curves recorded on different sensor GNR's in presence of the same microemulsion.

\begin{figure}[htbp]
    \centering
    \includegraphics{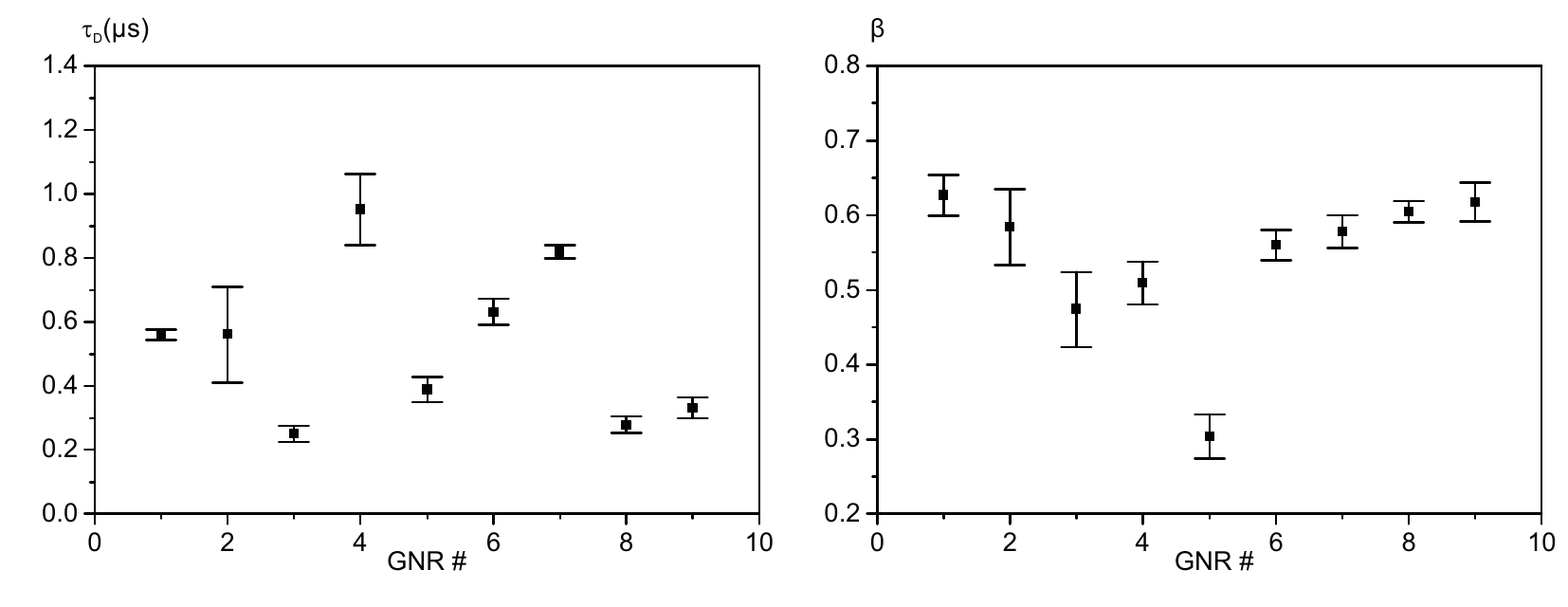}
    \caption{Averaged stretched-exponential fit parameters for decay time (left) and stretch exponent (right) obtained from $G_{100}$ curves that measured on different GNRs in the presence of microemulsion nanodroplets. Error bars indicate the standard deviation.}
    \label{fig:S2}
\end{figure}

\begin{figure}[htbp]
    \centering
    \includegraphics{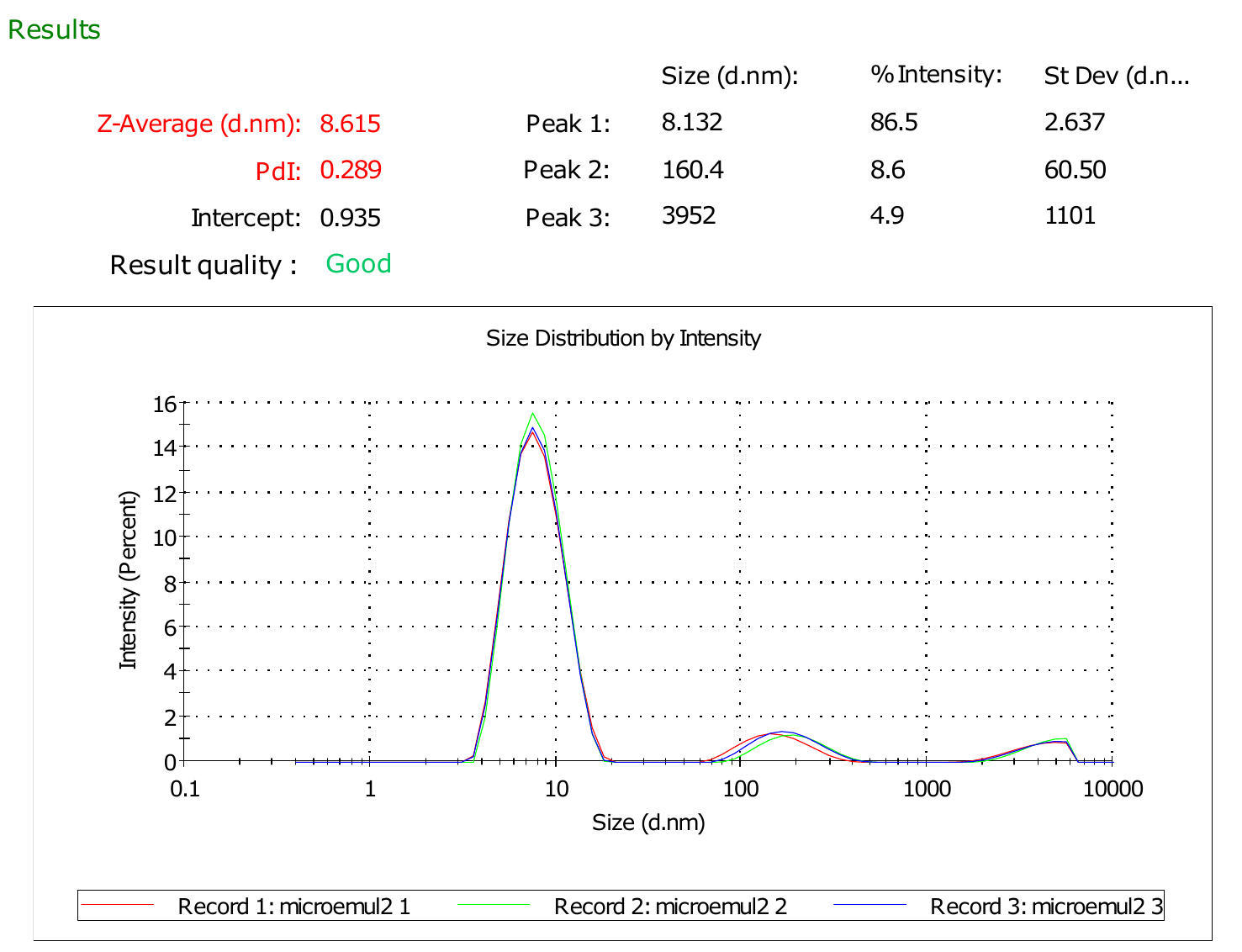}
    \caption{Size distribution of microemulsion nanodroplets measured via dynamic light scattering (Malvern Panalytical Zetasizer). Table values from left to right correspond to peak centers (particle size in nanometers), relative intensity attributed to the peak (\% Intensity: This is a measure of particle concentration and scattering cross section/size i.e. the smaller the particles the higher the concentration for a constant percentage of intensity) and the width of the peak (standard deviation) i.e. width of the size distribution - the narrower the distribution the more monodisperse is the sample. Peak $1$ is caused by the microemulsion nanodroplets. Peak $2$ and $3$ are contaminants i.e. bigger particles with in comparison to the nanodroplets significantly lower concentrations.}
    \label{fig:S3}
\end{figure}

We find that the $\tau_D$ values strongly vary between different GNR's (Fig. S2). This variation could arise from surface features on, as well as size and shape differences between, individual GNRs. Nonetheless most of the GNR's exhibit comparably similar stretching exponents in the presence of our microemulsion. Note that example autocorrelations for this are shown in Fig. S4A.
The size distribution of the microemulsion was measured via dynamic light scattering (DLS, Malvern Zetaziser) and is shown in figure S3.

\subsection{S4: Autocorrelation}
\textbf{Computation using Wiener–Khinchin theorem:}

We compute autocorrelation curves $G(\tau)$ from our experimental intensity traces $\mathbf{I}=[I_0,...,I_N]$ using the Wiener–Khinchin theorem. Specifically we compute: 
\begin{equation}
   \frac{\mathfrak{F}^{-1}\left(\mathfrak{F}( \mathbf{\tilde{I}})
   \mathfrak{F}^*(\mathbf{\tilde{I}})\right)}{\sigma^2(\mathbf{I})},
\end{equation}
where $\mathbf{\tilde{I}}=\mathbf{I}-\langle \mathbf{I} \rangle$, $\mathfrak{F}$ denotes the Fast Fourier Transform (FFT), $(^*)$ indicates the complex conjugate, $\mathfrak{F}^{-1}$ the inverse FFT and $\sigma^2$ the variance.

\textbf{Autocorrelation contrast:}

Regarding the concentration dependence of our autocorrelation curves we find that the contrast increases with increasing concentration (see Fig. S4B).
\begin{figure}[htbp]
    \centering
    \includegraphics{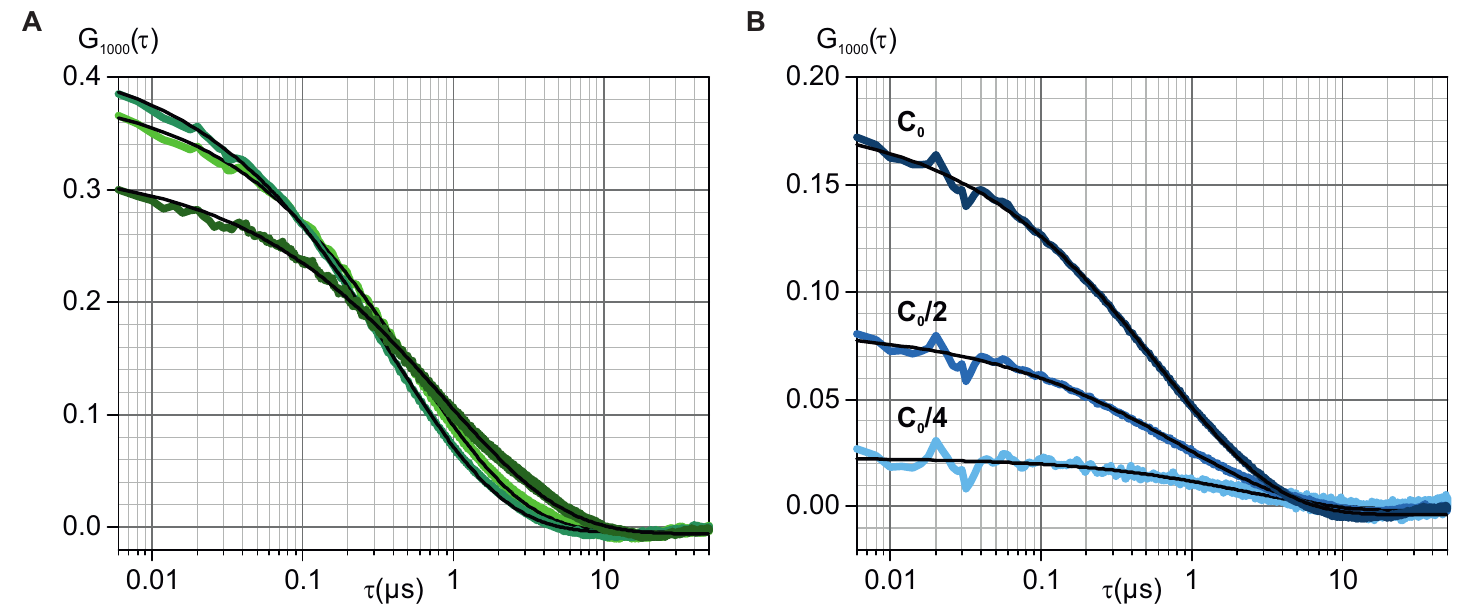}
    \caption{Microemulsion: Comparison between rods and concentration dependence. The averaged autocorrelation curves (green) shown in panel \textbf{A} together with their respective fits (black) were recorded with independent sensor GNRs and microemulsion samples. Panel \textbf{B} shows the evolution of averaged autocorrelation curves (blue) and fits (black) as the microemulsion sample is diluted stepwise (dark blue to light blue) to a quarter of its initial concentration. The features between $20$ and $30$\,ns are artefacts from the systems electronic response.}
    \label{fig:S4}
\end{figure}

We defined our autocorrelation function as:
\begin{equation}
G(\tau)=\frac{\langle \tilde{I}(t+\tau)\tilde{I}(t)\rangle }{\langle \tilde{I}^2 \rangle},
\end{equation}
where $\tilde{I}(t)=I(t)-\langle I\rangle$.
We can rewrite this intensity fluctuation as a sum of fast fluctuations of the detector itself, which we approximate as white random noise $\sigma(t)$, and signal fluctuations $p(t)$, which are correlated between consecutive measurements, i.e., $\tilde{I}(t)=(\sigma (t)+p(t))$. We assume both fluctuations to be centered in zero: $\langle p(t)\rangle=0$ and $\langle\sigma(t)\rangle=0$. Moreover, we suppose these fluctuations to be uncorrelated, i.e., $\langle \sigma(t) p(t+\tau) \rangle_t=0$ for all $\tau$.
As we will use Fourier transformations to calculate the correlation function, we discretize our measurement trace. The detector's white noise will end up in the first autocorrelation channel $\langle \sigma(t)\sigma(t) \rangle_t=\sigma_0^2$, whereas $\langle\sigma(t)\sigma(t+\tau)\rangle_t=0$ for all $\tau \neq  0$. Further $\langle p(t)\rangle=0$. So we can rewrite $G(\tau)$ as:
\begin{equation}
G(\tau)=\frac{\langle \sigma(t+\tau)\sigma(t)\rangle + \langle p(t+\tau)p(t)\rangle }{\sigma^2_0  + \langle p^2 \rangle}.
\end{equation}
Clearly $G(0)=1$, however for $\tau>0$:
\begin{equation}
G(\tau)=\frac{\langle p(t+\tau)p(t)\rangle }{\sigma^2_0  + \langle p^2 \rangle},
\end{equation}
Consequently we can rewrite as:
\begin{equation}
G(\tau)=\frac{\sigma_0^2}{\sigma^2_0  + \langle p^2 \rangle}\delta_{0\tau}+\frac{\langle p^2 \rangle }{\sigma_0^2 + \langle p^2 \rangle}\left(\frac{\langle p(t+\tau)p(t)\rangle }{\langle p^2 \rangle}\right),
\end{equation}
where $\delta$ denotes the Kronecker-Delta.  
Therefore, in the presence of white detector noise, we find that the amplitude of the correlation is reduced by the factor $G_{exp}=\frac{\langle p^2 \rangle }{\sigma_0^2 + \langle p^2 \rangle}$.

We can now identify two factors that will contribute to the contrast. Firstly there is a maximum amplitude of intensity change that can be caused by any single analyte. This maximum amplitude in relation to noise level gives the maximum contrast that a single perturbation can achieve. So this is essentially the influence of the tuning function and the analytes as well as the NR's properties. This is essentially due to the normalization to the total variance, and is apparent for analytes with weak perturbations i.e. if $\langle p^2_{max} \rangle < \sigma_0^2$. In this regime an increase in the rate of perturbations will yield an increase in contrast i.e. as $\langle p^2\rangle \approx n \langle p_{single}^2 \rangle $ until $\langle p^2 \rangle  \gg \sigma^2_0$. Once this limit is reached no further increase in contrast can appear. If single perturbations are already strong in comparison to the noise floor few or even a single perturbation will cause significant contrast (compare Fig, S8 C and E). The measurements shown in Fig. S4B fall into the regime of weak perturbations as they exhibit an initial contrast $<0.2$ at the highest concentration. Reduction of nanodroplet concentration during these measurements ensured that we remain in this regime and the contrast exhibits the expected decrease as the concentration (i.e. the rate of perturbations) is reduced.

\subsection{S5: Debye length}
The Debye length $\lambda_d$ was calculated according to:
\begin{equation}
\lambda_d=\sqrt{\frac{\epsilon_r \epsilon_0 k_b T}{2e^2IN_A}},   
\end{equation}
where $\epsilon_r$ ($\epsilon_0$) denote the medium's relative permittivity (the vacuum permittivity), $k_B$ is the Boltzmann constant, $e$ the elementary charge, $N_A$ is Avogadro's number and $I$ denotes the ionic strength of the electrolyte: 
\begin{equation}
I=\frac{1}{2}\sum_i  c_i z_i^2,    
\end{equation}
where $i$ is an index for each ionic species in the solution, $c_i$ are the corresponding concentrations (note with above formula this has to be entered in units of mole per cubic meter) and $z_i$ is the corresponding number of elementary charges\cite{Israelachvili2011}.
For a salt composed of monovalent ions (i.e. NaCl) the ionic strength equals the concentration of the electrolyte. 

\subsection{S6: Single-event Detection/Analysis}
We use an AC-coupled detector for our measurements. In consequence the global mean of the recorded signal equals $0$. In the absence of external perturbations deviations from this value are due to electronic and shot noise, which give rise to the background standard deviation of $\sigma_{BG}$. We assume this background noise is normally distributed. If a limited number $N$  of intensity samples $I_i$ is observed this background standard deviation, even in absence of additional perturbations can deviate from its global value. The magnitude of this unperturbed deviation scales $\propto\sigma/\sqrt{N}$. Intensity perturbations due to analyte particles will give rise to fluctuations in addition to this background. In order to find these perturbations we search for intervals where the local mean $\langle I \rangle_N$, where $N$ indicates the number of points used, exceeds the global deviation by a threshold factor $M_T$ taking into account the scaling of fluctuations due to limited sample size.
This is expressed by the following inequality: 
\begin{equation}
  \vert \langle I \rangle_N \vert -\sigma_{BG}\geq M_T \frac{\sigma_{BG}}{\sqrt{N}}.
\end{equation}

\begin{figure}[hbtp]
    \centering
    \includegraphics{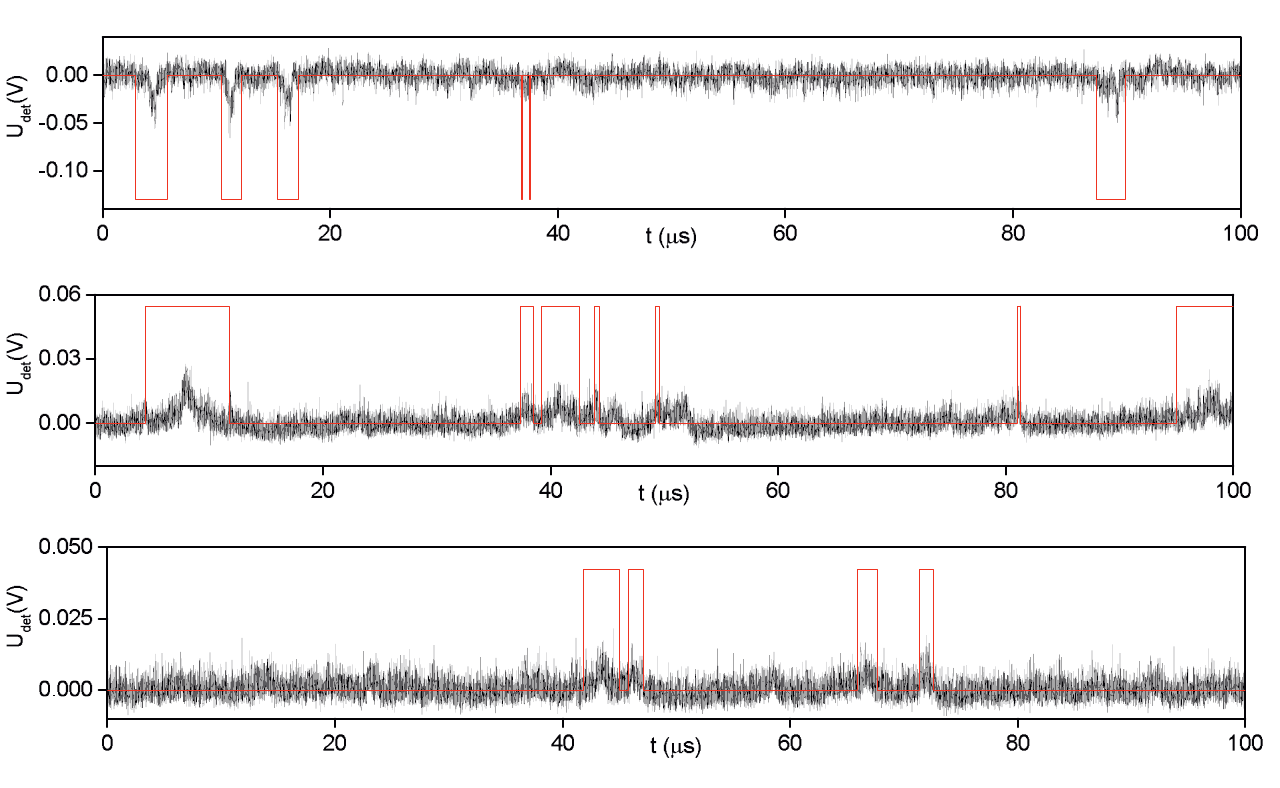}
    \caption{Example traces showing how the algorithms functions. Experimentally obtained traces $\mathbf{I}$ are black. Red lines are used to indicate where the algorithm finds events.}
    \label{fig:S5}
\end{figure}

As the length of events and their location is unknown we check this inequality for trace intervals with varying length of $L=32$, $64$, $128$, $256$, $512$, $1024$ and $2048$ points and inter-interval spacing of by $\Delta i= L/16$\, points. Intervals that fulfill above inequality are marked as events and used for further analysis. Experimentally obtained example traces in the presence of $5$\,nm diameter GNPs and the corresponding intervals marked as events are shown in Fig. S5.

\textbf{Negative Controls:}
We have computed traces of Gaussian white noise in order to determine the number of false positives for different values of $M_{T}$. Specifically we have run the simulations for a total of $10^8$ points for integer $M_{T}$ values ranging from $0$ to $5$, which yielded no events for either value. Thus we ran $5$ continuous tests for $M_{T}=0$ which were stopped as soon as one false positive was found in a simulation step consisting of $10^5$ points. This was the case after a total of $272.4\cdot10^6$, $2.2\cdot10^6$, $100.3\cdot10^6$, $379.1\cdot10^6$ and $3.9\cdot10^6$ points. Thus it is statistically unlikely that the above algorithm reports false positives even for the lowest threshold setting $M_{T}=0$.  
For our proof of principle experiments we have chosen a relatively hard value of $M_{T}=7$ to account for variations of intensity due to possible experimental imperfections like mechanical drifts, vibrations or fluctuations in laser intensity. 

\begin{figure}[htbp]
    \centering
    \includegraphics{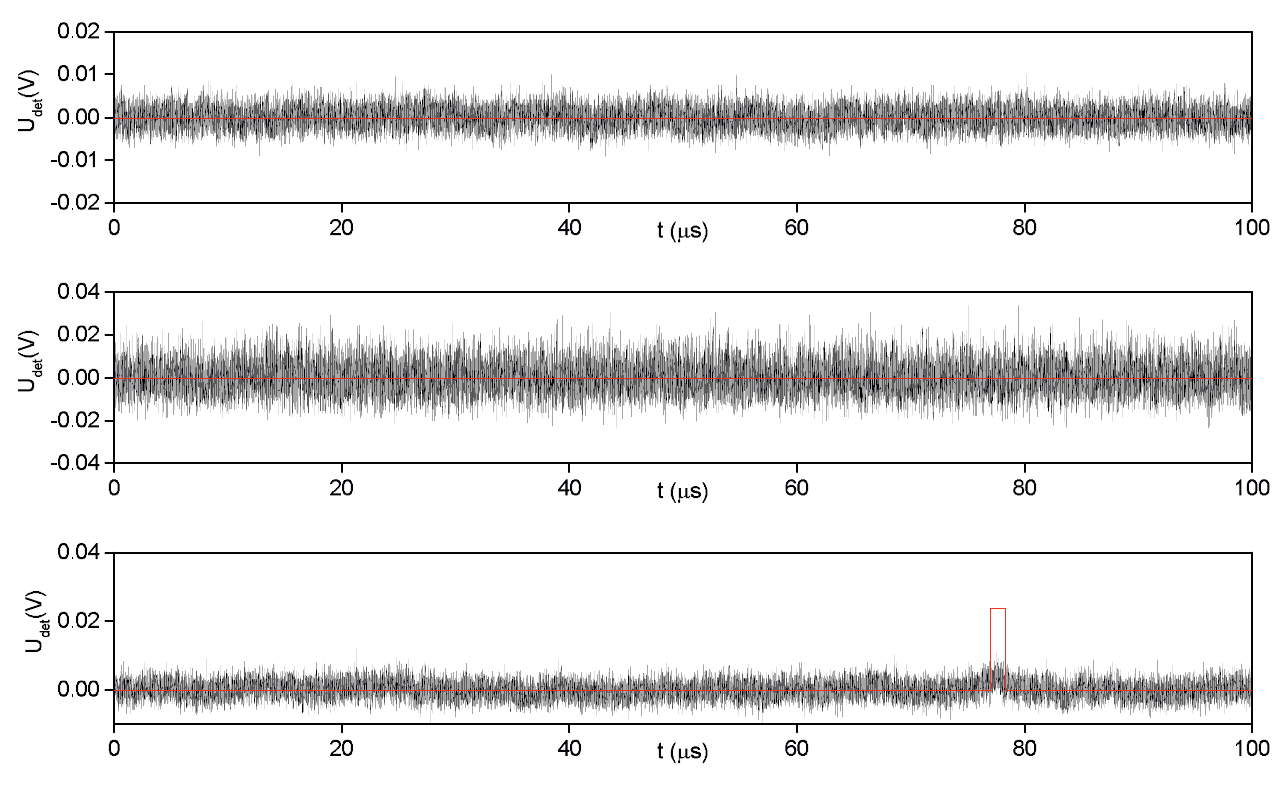}
    \caption{Experimentally obtained example traces (black) and event finding algorithm (red). The top and bottom traces were recorded in the absence of analyte GNPs. The middle trace was recorded in presence of GNPs (diameter: $5$\,nm) with $8$\,nM concentration in the absence of additional electrolyte (Milli-Q water). The bottom trace shows the strongest of the few events found in absence of analyte particles, possibly caused by solution impurities.}
    \label{fig:S5}
\end{figure}

We have further recorded traces in the absence of analyte particles (see Fig. S6 top and bottom) for which we found a total of $3$ events during a recording time of $0.5$\,s ($2.5\cdot10^8$ points). These very few events could be due to impurities (i.e. non-analyte particles) in the solution. One of these $3$ events, the one with the highest amplitude, is shown in figure S6 (bottom). In comparison we find rates of $\approx 5\cdot 10^4$ events per second in the presence of analyte GNPs ($5$\,nm diameter) with a concentration of $36$\,nM.
We have also recorded traces (Fig.\,S6 middle) in the presence of $5$\,nm diameter GNPs (concentration $8$\,nM) but in the absence of electrolyte. Under this condition the repulsive Coulomb interaction between analyte GNPs and the observed nanorod prevents the GNPs from entering the NR's near field and the algorithm recognizes no events during a recording time of $1$s ($5\cdot10^8$ points). 
We only considered negative controls on nanorods and that also yielded positive results for single $5$\,nm GNP detection.

\textbf{Determination of Event Properties:}
The above algorithm, due to the limited set of interval sizes, will bias the length of events. Moreover perturbations of higher amplitudes are likely to be associated with longer intervals which can exceed the length of the actual perturbation. Nonetheless above algorithm provides us with zones of interest for further analysis. 
We will firstly divide events into two categories. 
For this we run a median filter with a length of $L_{MF}=L_{EV}/20$ limited to a maximum of $L_{MF,max}=100$ over our dataset $\mathbf{I}$. This removes noise from the dataset, however, also rejects short bursts. We then check if the dynamic range i.e. the difference between the maximum and the minimum of the median filtered event-dataset $\mathbf{Y}$ exceeds at least two times the standard deviation: $|Y_{max}-Y_{min}|\geq2\sigma_{BG}$. 
If this condition is not fulfilled the event falls into category I and if it is fulfilled the event falls into category II. 
\begin{figure}[htbp]
    \centering
    \includegraphics{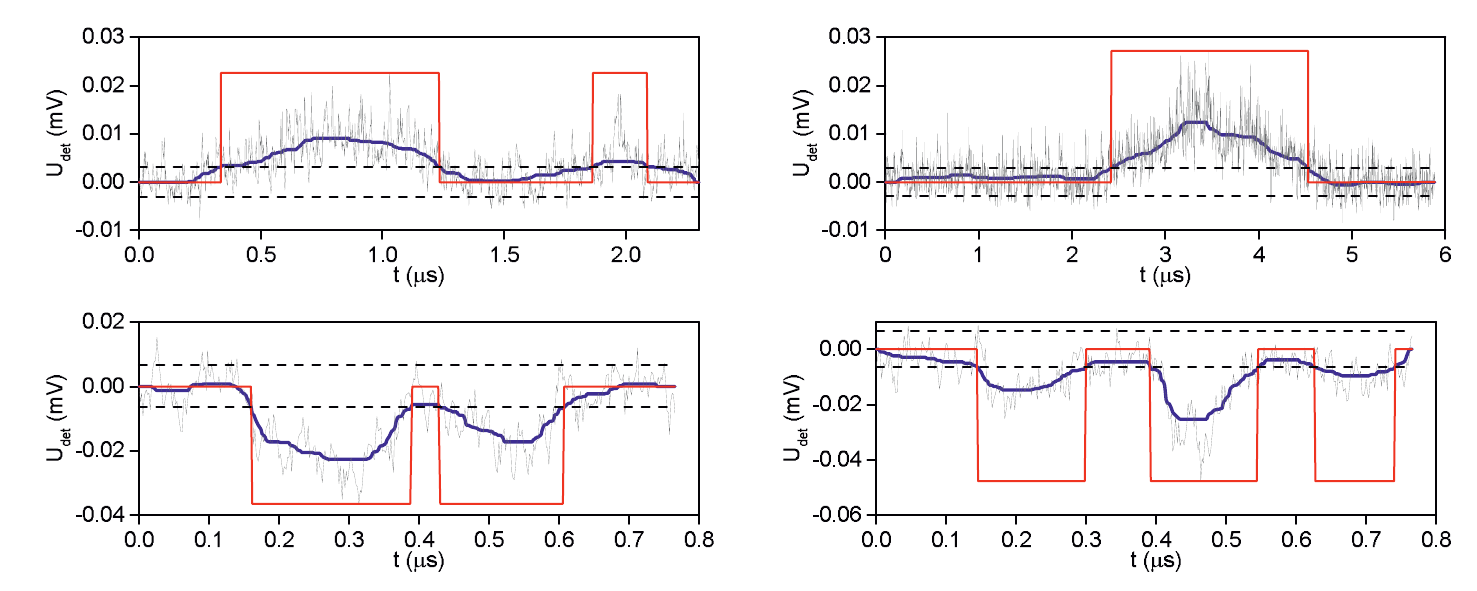}
    \caption{Example plots illustrating the analysis of intervals found by the algorithm (14). Raw data $\mathbf{I}$ is depicted grey. Dark blue lines indicate the median filtered data $\mathbf{Y}$. Dashed lines indicate the threshold $\pm\sigma_{BG}$ red boxes mark single events and enclose the data points used for further analysis.}
    \label{fig:S7}
\end{figure}
Events of category I are only analyzed with respect to their mean amplitude and the length, begin and end of the interval determined via the event finding algorithm will be used as their duration, start- and end-points for further analysis.  
Events of category II possess enough dynamic range for further analysis. Specifically we determine the start $i_s$ and duration $\Delta i$ of individual events by using all consecutive points for which $|Y_i|\geq\sigma_{BG}$. If multiple separated intervals fulfill this condition they are considered as separate bursts for further analysis. Example intervals, found by the algorithm eq.\,(14), demonstrating this process are depicted in figure S7.
Burst attributes such as mean amplitude $\langle I\rangle$, integrated intensity, maximum amplitudes, variance \[\sigma^2 = \left(\sum_{k=i_s}^{i_s+\Delta i-1}(I_i-\langle I \rangle)^2\right)/\Delta i\]  and higher statistical moments such as skewness and kurtosis are then determined using the unfiltered data points $I_i$ corresponding to the intervals with $|Y_i|\geq\sigma_{BG}$.

\subsection{S7: Distributions of inter-event durations:}

\begin{figure}[htbp]
    \centering
    \includegraphics{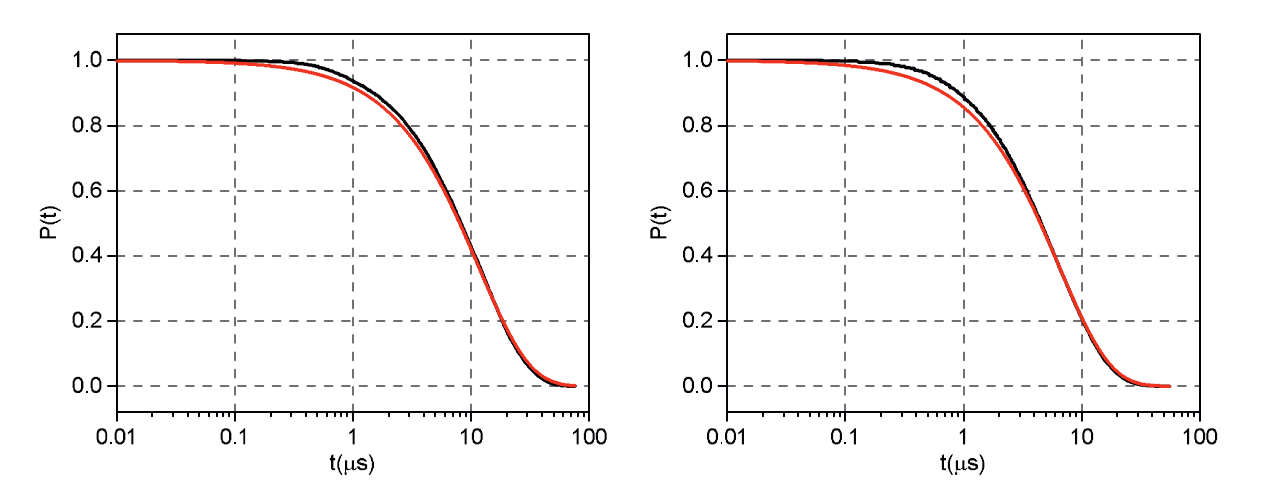}
    \caption{Typical probability distributions for the durations between events caused by $5$\,nm diameter GNPs. Left(Right) Panel for excitation wavelength blue (red) detuned from the sensor NR's LSPR. Black Curves are experimental values. Red curves are fits to single exponential decays.}
    \label{fig:S6}
\end{figure}
The times between perturbations caused by single particles should be poissonian distributed i.e. the probability $P_0(t)$ for finding zero events in an interval of length $t$ should follow : $P_0(t)=e^{-t/\tau}$ where the decay time $\tau$ is the only parameter.
We obtain our inter event time probability distributions without binning to avoid the errors commonly associated when fitting exponential decays to histograms as in that case the choice of bin size and starts can heavily influence the resulting fits. 
Specifically we use the method described by M.D.Baaske et al.\cite{Baaske2016}. We take into account that our traces are recorded with dead times between them - so we cannot determine the time after the last event of each trace. This also restricts the maximum duration between events to less than the length of individual traces i.e. $0.1$\,ms. Example probability distributions are depicted in Fig. S8. Generally we find our distributions to  be in good agreement with the expected poissonian distributions (single exponential decay). 
The deviations of the experimental distributions from the fits (fig. S8) can be caused by fluctuations of the event rate over time. The origin of such fluctuations can be bulk temperature fluctuations which alter the diffusion constants or any movement of the solution altering the local analyte density around the sensor NP over time.

\subsection{S8: Simulation of signals via 3D-diffusion}
We perform Monte-Carlo simulations of particles freely diffusing in a box containing a single nanorod. We restrict our simulations to one particle meaning that simulations which compare different particles are performed by simulating multiple separate boxes rather than multiple particles in the same box. 
Coordinates $x_i$ ($i=1,2,3$) are obtained iteratively with time-steps of $\tau=2$\,ns i.e. $x_i(t_{k+1})=x_i(t_{k})+\sqrt{D\tau}q_i$, where the $q_i$ ($i=1,2,3$) are normally distributed random variables with a standard deviation of $1$.   
The interaction between the particle the NR and the walls is restricted to reflection and if desired sticking. Whereas sticking occurs with a fixed probability when the particle hits the NR or the glass surface (wall of simulation box closest to the NR) and is modeled as poissonian process i.e. the likelihood to stick for a certain time $t_s$ follows an exponential decay $P(t_s)=e^{-t_s/\tau_s}$ where $\tau_s$ is the sticking decay time.
From the simulated coordinates we obtain simulated intensity changes using a sixth-order power law (dipole-induced dipole interaction):
\begin{equation}
I(r) = A_r\frac{1}{\left(1+r/L_d\right)^6},    
\end{equation}
where the decay length $L_d$ is the distance at which $I(L_d)=I(r=0)/64$ and $r=r_C-R$ with the particle's distance $r_C$ from the closest center of the NR's two hemispheres and the hemisphere's radius $R$. $A_r$ indicates the relative amplitude at $r=0$. We model the particle as solid sphere in consequence the minimum distance $r_{min}=r_p$, where $r_p$ is the particle's radius.  
For the determination of $I(r)$ we assumed point like dipoles. The diffusion process, however, is modeled with hard spheres. Therefore the minimum distance to the NR's surface equals the particles radius and maximum amplitudes are: $I_{max}=I(r_{p})$. 
\begin{figure}[p]
    \centering
    \includegraphics{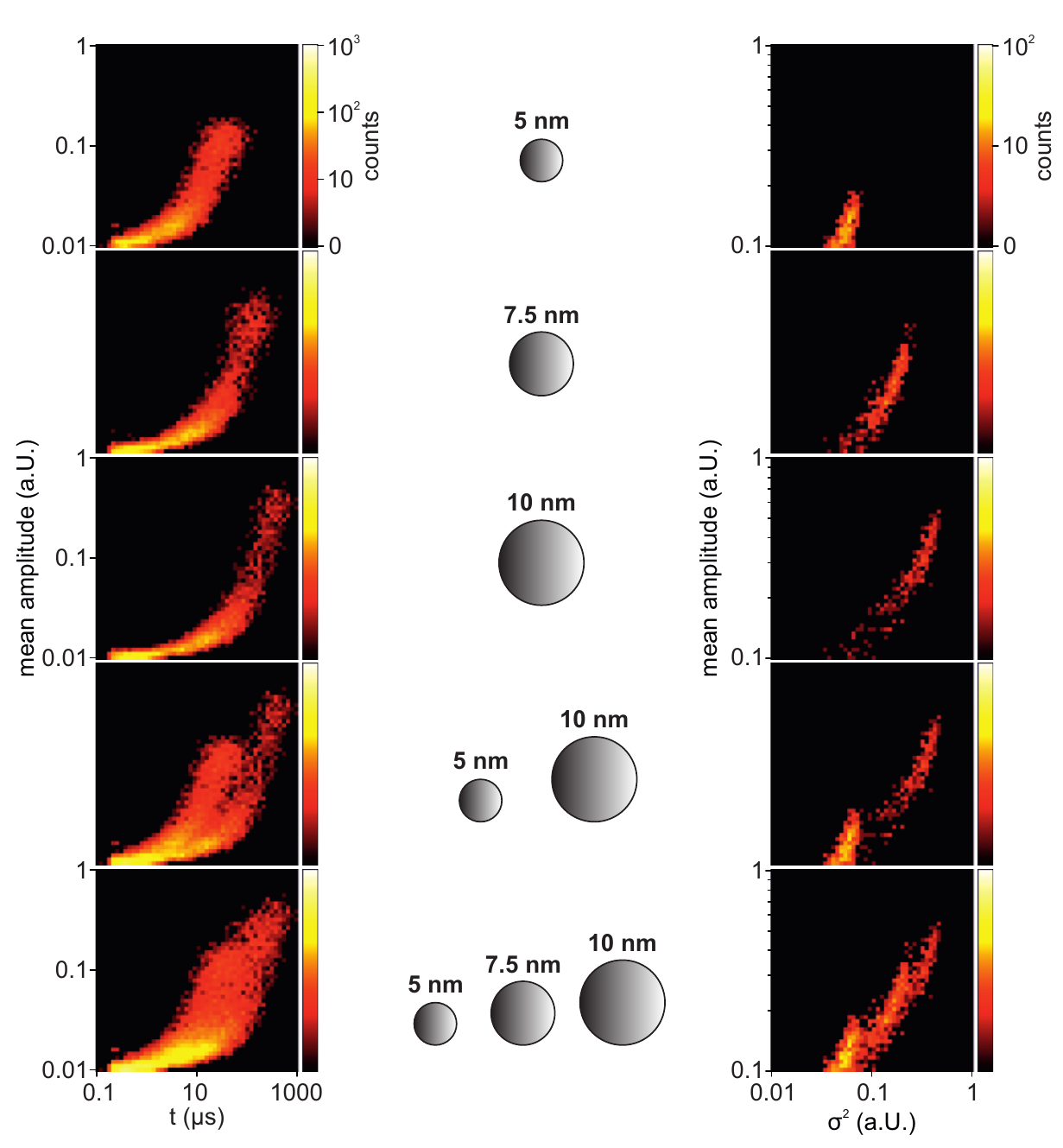}
    \caption{Simulation of single-particle detection: The panels on the left side show mean amplitude vs. duration distributions. The panels on the right side show a zoom-in on areas of interest in the mean amplitude vs. variance ($\sigma^2$) distributions. The middle panel indicates the diameter of the simulated particles used for the adjacent distributions. Distributions with more than one particle size are cumulative displays of the respective single particle data sets (top 3). Relative amplitudes are scaled with the particles volume (i.e. $\propto$ polarizability): $A_r=1$, $3.375$ and $8$ for particle diameters of $5$, $7.5$ and $10$\,nm respectively. The noise level (white Gaussian noise) was fixed to $0.01$. Maximum signal-to-noise ratios are $\approx 50$, $120$ and $210$ for $5$ , $7.5$ and $10$\,nm diameter particles. All simulations where performed with $L_d=20$\,nm.}
    \label{fig:S9}
\end{figure}
Simulation results for particles with diameters of $5$, $7.5$ and $10$\,nm are shown in Fig. S9. The results show that the distinction of particle properties is essentially possible on a single event basis, especially if multiple dimensions of single-event properties and their relations are used. For example the two bottom panels (mean amplitude vs duration) on the left side of Fig. S9 show distinct and well separated zones for $5$ and $10$\, nm particles, however, these zones have quiet some overlap with the one obtained for $7.5$\,nm particles. Nonetheless these events can still be separated by size if the correlation between event mean amplitudes and variance, which exhibits distinct zones for all $3$ particle sizes (compare Fig.\, S9 bottom right) is used in addition. This shows that discrimination of particles sizes is possible, even for single events, as soon as multiple parameters are used.
Naturally this discrimination is restricted to events with higher amplitudes i.e. particles that come close to the NR's surface as they contain more information.

\subsection{S9: Additional Material}
\begin{figure}[hb]
    \centering
    \includegraphics[width=150mm]{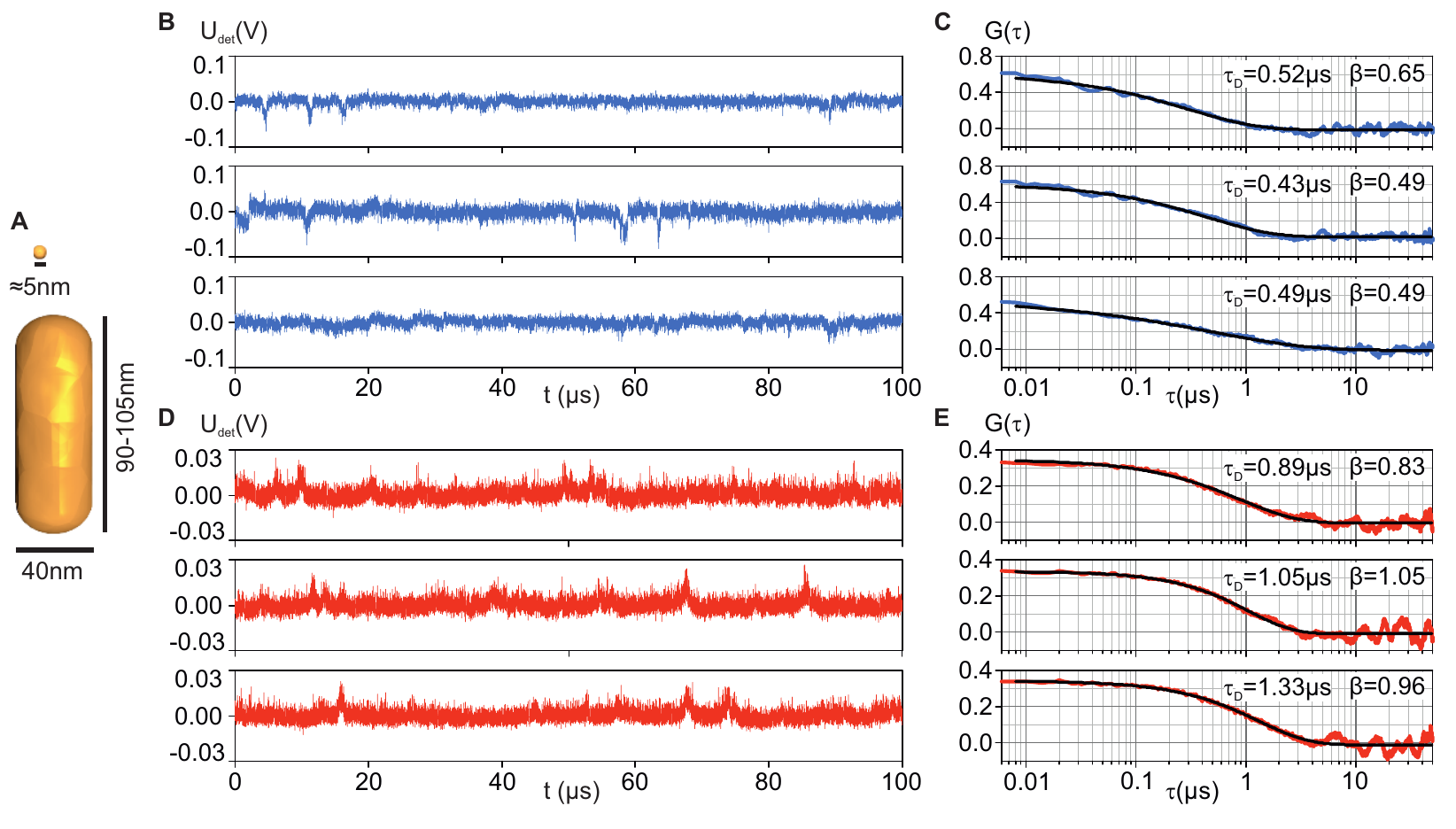}
    \caption{Single-particle detection. Similar to Fig. 3 of the main mansucript, Panel \textbf{A} shows the dimensions of the nanorods sensors and the analyte GNPs. Intensity traces are same as in Fig.\,3. and  Panels \textbf{C} and \textbf{E} show the corresponding autocorrelation curves (blue, red) and their respective stretched-exponential fits (black).}
    \label{fig:S8}
\end{figure}
\begin{figure}[ht]
    \centering
    \includegraphics{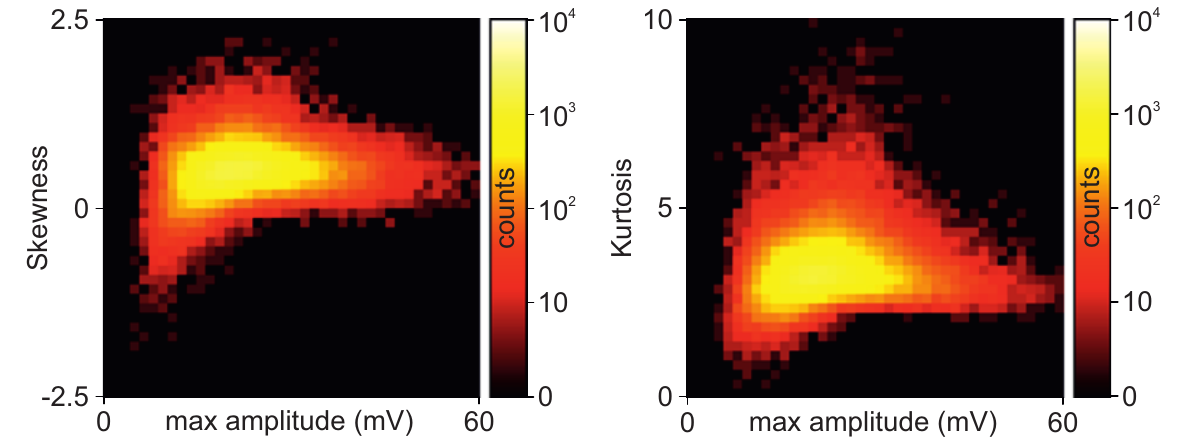}
    \caption{2D histograms of higher statistical moments (Skewness and Kurtosis) of single particle events vs. their maximum amplitudes. Values were obtained from the same data set represented in Fig, 3D and 3E i.e. perturbations are due to $5$\,nm diameter GNPs at $120$\,mM ionic strength.}
    \label{fig:S9}
\end{figure}

\bibliography{references}